\documentclass[twocolumn]{aastex631}

\graphicspath{{./}{figures/}}

\received{18 March 2022}
\accepted{10 May 2022}

\submitjournal{ApJ}

\shorttitle{SN~Ia properties from gamma-rays}
\shortauthors{Leising}

\begin{document}

\title{Deriving Thermonuclear Supernova Properties from Gamma-Ray Line Measurements}  

\correspondingauthor{Mark Leising}
\email{lmark@clemson.edu}

\author{M. D. Leising}
\affiliation{Clemson University \\
Department of Physics and Astronomy \\
Clemson, SC 29634-0978, USA}

\begin{abstract}
We illustrate methods for deriving properties of thermonuclear, or Type Ia, supernovae, including synthesized $^{56}$Ni mass, total ejecta mass, ejecta kinetic energy, and $^{56}$Ni  distribution in velocity, from gamma-ray line observations. We simulate data from a small number of published SN~Ia models for a  simple gamma-ray instrument, and measure their underlying properties from straightforward analyses. Assuming spherical symmetry and homologous expansion, we calculate exact line profiles for all $^{56}$Co and $^{56}$Ni lines at all times, requiring only the variation of mass density and $^{56}$Ni mass fraction with expansion velocity as input. By parameterizing these quantities, we  iterate the parameters to fit the simulated data. We fit the full profiles of multiple lines, or we integrate over the lines and fit line fluxes only versus time. Line profile fits are more robust, but in either case, we can recover accurately the values of the aforementioned properties of the models simulated, given sufficient signal-to-noise in the lines. A future gamma-ray mission with line sensitivity approaching 10$^{-6}$ photons cm$^{-2}$~s$^{-1}$ would measure these properties for many SN~Ia, and with unprecedented precision and accuracy for a few per year. Our analyses applied to the reported $^{56}$Co lines from SN 2014J favor a low $^{56}$Ni mass and low ejecta mass, relative to other estimates. 
\end{abstract}


\section{Introduction} \label{sec:intro}
Type Ia supernovae (SN~Ia) are understood as the thermonuclear explosions of white dwarf stars \citep{1960ApJ...132..565H,1984ApJ...286..644N}.
They are  essential tools in modern observational cosmology, because of their prominence in the upper rungs of the cosmic distance ladder  \citep{1998AJ....116.1009R,1999ApJ...517..565P}, and important contributors to nucleosynthesis of iron peak elements and energy input into the interstellar medium. Many are used as easily calibrated standard candles, however, neither their progenitor systems nor the processes of their nuclear ignition and subsequent burning are well constrained \citep{2019NatAs...3..706J}.

Great effort is applied to improving the determination of absolute magnitudes in spectral bands. 
It would  prove useful to measure basic physical properties of the explosions in order to better
understand their causes, and the relationships among their observed characteristics. Quantitative, model-independent
measurements have been elusive. Various timescales, such as light-curve widths, rise times, and decline rates are readily measured, but are somewhat removed from fundamental properties. Measurements of the mass of $^{56}$Ni, which dominates the power input to SN Ia ejecta and is an essential, but not unique, diagnostic of explosion models, have been most common \citep[e.g.,][]{2006A&A...450..241S}. Measurements of total ejecta masses and kinetic energies are more challenging, but would greatly help constrain the possible models for a given event, and ultimately lead to understanding of the progenitors of  SN~Ia and their variety \citep{2014MNRAS.445.2535S}. 

Most $^{56}$Ni mass measurements require several assumptions. For example, one can measure the flux in a few spectral bands at maximum light, derive the ``bolometric'' flux using  assumptions about the rest of the UVOIR spectrum, correct for extinction by dust, assuming supernova colors and interstellar dust properties, and employ a version of Arnett's rule where peak luminosity is proportional to radioactive decay power, and therefore infer an initial nickel mass \citep[e.g.,][]{2006A&A...460..793S}. 
Later nebular line measurements with modeling can provide independent $^{56}$Ni masses \citep{2015MNRAS.450.2631M}.
These depend on assumptions or models of ionization state, gamma-ray and positron energy deposition, and line identifications. 
\citet{2015MNRAS.454.3816C} derive $^{56}$Ni masses for a few tens of supernovae from cobalt line emission. \citet{2016A&A...588A..84D} avoid assumptions about interstellar dust extinction, using the correlation  of the time of the second infrared light curve maximum with the peak bolometric luminosity.

SN~Ia ejecta masses can be estimated from $^{56}$Ni mass measurements as above along with some modeling to explain the partial deposition of the radioactive decay power at later time. The effective mass overlying the radioactivity can be derived from the time  (often called t$_{0}$) at which the e$^{-1}$ of the radioactive decay power escapes, as inferred from the reconstructed UVOIR bolometric luminosity \citep{2006A&A...450..241S}. A single effective opacity is assumed for deposition of energy by multiple Compton scatters of gamma-ray photons. This simple model, equivalent to a single attenuating shell, for estimating the ejecta mass is sometimes modified by describing the distribution of $^{56}$Ni in the ejecta with another parameter \citep{1999astro.ph..7015J}.
\citet{2014MNRAS.440.1498S} add to this more constraints from SN~Ia models and conservation of energy (white dwarf binding, nuclear, and kinetic energies) to refine ejecta mass estimates. They also find a correlation between light curve stretch and their ejecta mass estimates, which they and others exploit \citep{2015MNRAS.454.3816C,2019MNRAS.483..628S}. \citet{2019MNRAS.484.3941W} turn this around by determining the parameter t$_{0}$ from the bolometric light curve shape in a distance independent manner, and then deriving the the $^{56}$Ni mass from the luminosity near time t$_{0}$. They also strongly caution against using t$_{0}$ to determine total ejecta mass, pointing to models with different masses but similar t$_{0}$'s. Typically, these methods yield estimates of $^{56}$Ni and ejecta masses to precision of 0.1--0.2 M$_\odot$, but systematic uncertainties often dominate the statistical ones \citep{2014MNRAS.445.2535S,2015MNRAS.454.3816C}.

\citet{1969ApJ...155...75C} introduced the idea of  the utility of gamma-ray line observations for understanding supernovae. Their most prominent targets were $^{56}$Ni (t$_{1/2}$= 6.1~d) with lines at 158 and 812 keV, and 56Co (t$_{1/2}$= 77~d) with lines at 847 and 1238 keV, among others. Several of their predictions have been realized, but they were overly optimistic in their estimates of the frequency of significant detections of gamma-ray lines from supernovae. Only one core-collapse supernova, SN 1987A \citep{1988Natur.331..416M,1989Natur.339..122T}, 
and one thermonuclear supernova \citep{2015ApJ...812...62C,2015A&A...574A..72D}, 
have been detected in $^{56}$Co decay gamma-ray lines. Analyses of these SN~Ia data, as well as earlier upper limits, have mostly consisted of comparing the data with line flux histories from a number of previously calculated SN~Ia models.

In contrast to the assumptions and modeling needed to interpret UVOIR observations, using gamma-ray line measurements is straightforward, although measuring them is challenging. Escaping line photons are by definition unscattered, and they see only the total column density of electrons, bound and free alike, from the point of emission to the surface. This depth in the homologous expansion has a simple time dependence, as does the source decay.  The line opacity is independent of time and position in the ejecta, assuming the electron mole fraction, Y$_{e}$, varies little throughout the ejecta. If line fluxes can be measured to late times, they give very directly, with a small correction for the few still scattered photons, the remaining $^{56}$Co mass and thus the initial $^{56}$Ni mass. Interstellar and circumstellar dust are transparent to  gamma rays, and therefore irrelevant. The main limitation to using gamma rays to study SN~Ia is photon counting statistics, which will limit their usefulness to relatively local supernovae for the foreseeable future.  

If we could derive, for example, $^{56}$Ni and ejecta masses from gamma-ray observations with comparable or better precision than those above, and with small systematic uncertainties, the constraints on SN~Ia models would be significant. Ejecta masses clearly near the Chandrasekhar mass, or distinctly different, will point directly toward progenitor systems. $^{56}$Ni masses determined with high precision could aid in the challenging problem of UVOIR radiative transfer in SN~Ia. Measurements of the velocity distribution of $^{56}$Ni in the ejecta will help clarify the nuclear burning conditions. 

Here we focus only on  the nuclear lines. Many of the scattered photons escape in a continuum, with steps at the lines, down to just below 100 keV, where photoelectric absorption truncates the spectrum. There is information in the continuum, but its transfer is not as straightforward as that we describe herein for the lines, and after a few months the lines dominate the spectrum. Also, elimination or subtraction of the large backgrounds of MeV gamma-ray instruments is notoriously difficult. Even if some residual background remains, often the distinctive line shapes can be used to extract the source line photons from the residual spectra. We illustrate a method for extracting supernova density and $^{56}$Ni abundance profiles from gamma-ray line measurements. Integrating over these, we get $^{56}$Ni masses, ejecta masses, and kinetic energies, which turn out to be more robustly determined than the profiles themselves. 

\section{Simulated Data}
We create simulated gamma-ray data from a small number of representative numerical models of normal SN~Ia. Our goal is not to test all plausible models, but to determine the precision we can achieve for the basic SN~Ia properties. As we expect to determine $^{56}$Ni mass to good precision, we do not choose models that span a wide range in it, but as total ejected mass, M$_{ej}$, is more diagnostic of progenitors and more difficult to measure, we choose models that span the range of typical values for three progenitor classes. We use the sub-Chandrasekhar-mass model, HeD8  \citep{1996ApJ...457..500H}, the delayed detonation model, 7p0z22d20\_27, with central density 2.0$\times 10^{9}$ g~cm$^{-3}$ \citep{2015ApJ...806..107D}, and a spherically symmetric approximation to the massive white dwarf merger model of 
\citet{2012ApJ...747L..10P}, MWD. In order to test the gamma-ray line sensitivity to this effect, we also add a variation of the aforementioned delayed detonation model, 7p0z22d07\_27, which has a lower central density, 0.7$\times 10^{9}$ g~cm$^{-3}$ \citep{2015ApJ...806..107D}, resulting in less electron capture and more slow, central $^{56}$Ni.
\begin{table*}[htbp]
\centering
\caption{ Supernova Models Simulated  \vspace{3mm}}
\begin{tabular}{lccc}
\tableline \\[-4.5mm] \tableline
Model 	& M$_{ej}$ & M($^{56}$Ni)	& Kinetic Energy \\
 & M$_{\odot}$ & M$_{\odot}$ & 10$^{51}$ erg  \\
\tableline 
Hed8	& 0.96 & 	0.51 &	1.03   \\
7p0z22d20\_27	& 1.37 & 	0.65	& 1.35 \\
7p0z22d07\_27	& 1.37	& 0.74	& 1.35 \\
MWD	& 1.94	& 0.61	& 1.68 \\
\tableline 
\end{tabular}
\label{mods}
\end{table*}%

For each model, we use Monte Carlo calculations of the gamma-ray spectrum (generously provided by Peter H\"oflich) at twenty times, from 5 to 400 days post-explosion. We  ``measure'' each spectrum with a simplified gamma-ray detector. The instrument is defined by its narrow line sensitivity at 847 keV in a 10$^{6}$ second observation, and by its energy resolution, $\Delta$E/E, at 847 keV. A supernova distance is specified, the spectrum is scaled to that distance, rebinned to several bins per energy resolution element, and Gaussian random noise is added to achieve the specified sensitivity over an instrument energy resolution element. We then analyze each spectrum, either fitting it with model line profiles as described below, or, if analyzing total line fluxes only, we extract the flux of each line of interest. To do the latter, we fit a narrow range around each line with a Gaussian line profile plus a continuum described by a flat step function at the Gaussian central energy. The step is intended to remove the Compton continuum from scattering with the supernova ejecta, which appears below each line. The $^{56}$Co lines are sufficiently spaced in energy to fit each separately, but for early times, the 812 keV $^{56}$Ni and 847 keV $^{56}$Co lines overlap. With better energy resolution, we fit these as two Gaussian lines, while for poorer resolution, we fit both together as a single Gaussian, and apportion the flux to the two lines in the known time-dependent ratio. While this process is not perfect, we check it against the known line-only fluxes and profiles from our calculations (see Appendix.) In all cases, the extracted lines are consistent with the known values for these models within the statistical uncertainties. For real supernovae, in particular with high precision measurements, careful data analysis will be needed to extract just the lines.

Without being more specific about the instrument we are using, we make simple assumptions. We take the energy resolution, $\Delta E$ to vary as the energy $E^{1/2}$. Given that, if the background spectrum, which typically dominates counting statistics at these energies, is flat versus energy, the minimum detectable line flux would increase as E$^{1/4}$. Typically, background rates fall with energy in this range, so we simply take the line sensitivity to be the same for all lines. We analyze the resulting noisy simulated gamma-ray data to extract supernova properties. We first study the line fluxes, and defer the line profiles to Section~\ref{sec:profiles}.

\section{Single Scattering Shell}\label{sss}

A measurement closely related to t$_0$ discussed above is the time at which an escaping gamma-ray line luminosity is e$^{-1}$ of that emitted. The gamma-ray measured quantity, which we will here call t$_\gamma$, is more easily interpreted. In the approximation of a single overlying shell of material, a measurement of t$_{\gamma}$ provides the number of overlying electrons per unit area at that time, and therefore at all other times, assuming constant expansion. For high-precision line flux measurements, the deviation from a single  scattering shell will be apparent, because the $^{56}$Ni, itself an important fraction of total scattering medium, is found at a range of depths. Multiple line fluxes can  easily be fit simultaneously, with escaping fluxes differing due to only the  known branching ratios and total angle-integrated Klein-Nishina scattering cross-sections.

In the approximation of a single overlying spherical shell, only two quantities are required to describe all line fluxes at all times: the line luminosities, parameterized as initial M($^{56}$Ni), and the effective optical depth of the shell at some time. For constant expansion rate, the shell thickness varies as t$^{-2}$. We write the escape fraction of a gamma-ray line at energy E as
\begin{equation}
f_{esc}(E) = \exp(-\sigma_r(E)\frac{t_{\gamma}^{2}}{t^{2}}),
\end{equation}
where $\sigma_{r}(E)$ is the total Compton scattering cross-section at energy E in units of that at 847~keV, and t$_{\gamma}$, the second parameter in the light curve fits, is defined for 847 keV photons. This model was among those employed for the four $^{56}$Co gamma-ray line light curves of SN 1987A \citep{1990ApJ...357..638L}, but provided a poor fit to that massive core collapse supernova.

\subsection{M($^{56}$Ni) and t$_{\gamma}$ Fits}
To simulate a supernova measurement, we choose one of the above numerical models, define the instrument, choose a distance, and choose which lines to fit simultaneously. First we fit  line fluxes, so the instrument is defined by its sensitivity to these Doppler broadened lines. A nominal line sensitivity we use is a limiting flux at 3-$\sigma$ significance of $F_{3\sigma}$ = 1.0$\times10^{-6}$ cm$^{-2}$~s$^{-1}$ in a 10$^{6}$ s observing period. This is a quite ambitious sensitivity, far better than anything yet achieved. 
Note that a higher resolution detector with $\Delta E/E$ better than a few percent, will have its sensitivity to broad supernova lines degraded relative to its narrow line sensitivity, as more background counts are included under the broad line. 
\begin{figure*}[ht!]
\plotone{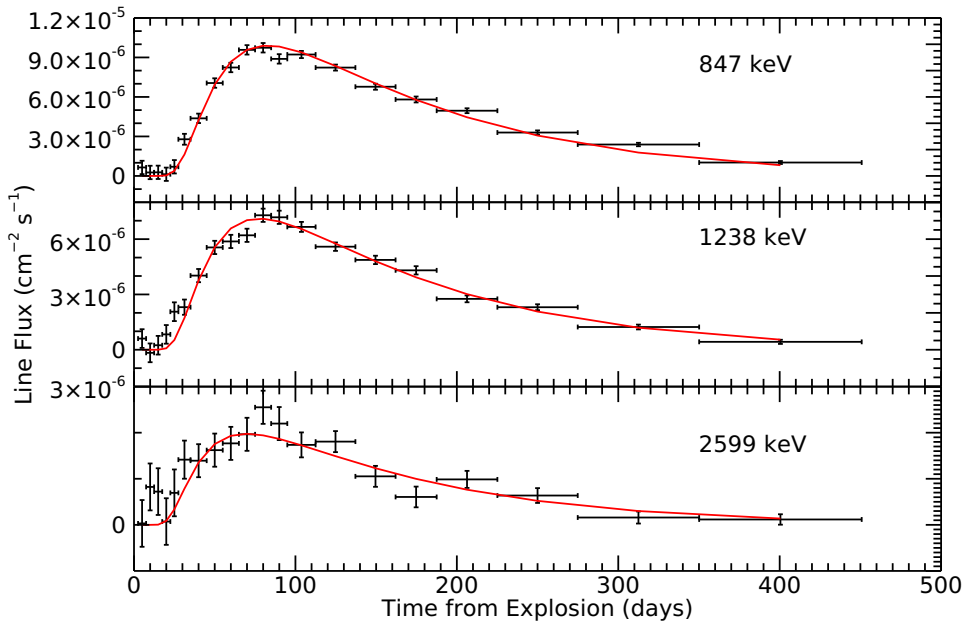}
\caption{An example of line fluxes from the simulated data of model 7p0z22d20\_27 fit with the two-parameter model. The distance is 20 Mpc, and the instrument line sensitivity is taken to be 1.0$\times10^{-6}$ cm$^{-2}$ s$^{-1}$ (3$\sigma$). The best fit parameters are M($^{56}$Ni) = 0.60$\pm$0.01 M$_\odot$, and t$_{\gamma}$=50.4$\pm$1.0 days. The $\chi^{2}$ per degree of freedom is 2.0 for 58 degrees of freedom, as is typical for fits with this distance and sensitivity combination. \label{fig1}}
\end{figure*}

For illustration, we  use a distance of 20~Mpc. This is approximately the distance within which the rate of SN~Ia is 1 y$^{-1}$, using the mean rate within a volume of radius 100 Mpc \citep{2010ApJ...723..329H,2011MNRAS.412.1473L,2020ApJ...904...35P}. The actual nearby supernova rate is probably higher. Based on SN~Ia discoveries in the past ten years \citep{2017ApJ...835...64G}     the SN~Ia rate within 20 Mpc is approximately 2 y$^{-1}$, ignoring incompleteness of the catalog and extinction corrections, effects which only increase the rate estimate.

We typically fit three lines at energies 847, 1238, and 2599 keV, which have branching ratios 1.0, 0.68, and 0.17, respectively \citep{1999TORI}. The other lines add relatively little signal, but, of course, they can be included as well. It is possible that some detector would have particularly low background at certain line energies, and it would be advantageous to include them. An example of such a fit is shown in Figure~\ref{fig1}. For these choices, with high-significance measurements, this simple model is not an adequate fit. The fit rises late and steeply, relative to the simulated data, and falls too steeply, undercutting the later points slightly. As mentioned above, this is easily understood, as in all the numerical models from which we draw simulated measurements, the $^{56}$Ni is found at a range of column depths. Actually, this is encouraging, suggesting more parameters are required, and more information can be gleaned from such measurements. Each simulated data point is derived from Monte Carlo calculations at a single time, but for our fits we assume the measurements are averaged over an interval as shown by the horizontal bars. The variation of the measurement precision with observation duration, as $\Delta$t$^{-1/2}$, for different measurement intervals $\Delta$t is apparent.

In order to see the collective information available to this method from a sensitive gamma-ray line instrument, we fit this two-parameter model to the fluxes of three lines (847, 1238, and 2599 keV) for each supernova in a randomly generated sample. We assume that SN~Ia occur uniformly in space to distance 100~Mpc, at a rate of 100~y$^{-1}$ over three years. As noted above, this probably undercounts very nearby events. We choose with equal probability among the four models used herein to generate the data for each event. Assuming again an instrument sensitivity of 1.0$\times10^{-6}$~cm$^{-2}$~s$^{-1}$ (3$\sigma$) for each of the three lines, the signal-to-noise ratio of the M($^{56}$Ni) measurement is above three for SN~Ia nearer than 90~Mpc, depending slightly on which model is simulated. Here and throughout this work, we do not include distance uncertainties in our error budget for derived quantities. 

\begin{figure}[ht!]
\plotone{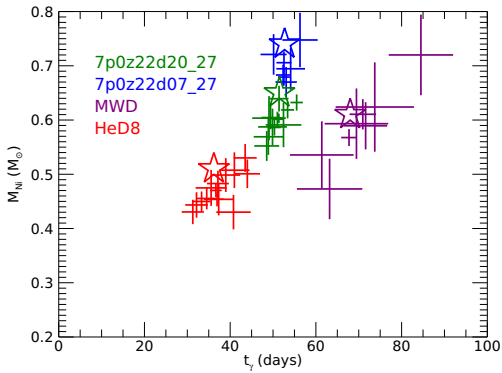}
\caption{Measured fit parameters in the single-shell model, for SN~Ia occurring within 50 Mpc over three years, for instrument line sensitivity 1.0$\times10^{-6}$~cm$^{-2}$~s$^{-1}$. Simulated supernova data is derived from the four models shown with equal probabilities. The stars show the $^{56}$Ni mass of each model from which the data were derived, and the t$_{\gamma}$ for a fit to a very nearby event.
\label{fig2}}
\end{figure}

Figure \ref{fig2} shows the measurements and uncertainties of the two fit parameters, color-coded by model used to generate the data. There we limit points shown to those nearer than 50~Mpc for clarity.  We see that the (uncorrected) M($^{56}$Ni) values are biased low relative to the true values (the stars). Also apparent are the correlations between the two parameters in the fitting process. The normalized fit correlation matrix typically has a value of 0.7. These four models are separated in these parameters, especially for nearer events, but it is likely that multiple numerical models among the many calculated overlap in this plane, even for some from different progenitors and explosion mechanisms. 

The inaccuracy of the single-shell approximation leads to a bias toward lower values in the measured $^{56}$Ni mass. The bias tends to be larger for lower values of t$_{\gamma}$, that is, for more distributed $^{56}$Ni. We can use this to improve the measurement of the $^{56}$Ni mass. We fit the deviations of the fitted $^{56}$Ni mass values from the true values vs. t$_{\gamma}$, for a large number of models from the literature \citep{2014ApJ...786..141T}.
We  find that multiplying the measured $^{56}$Ni mass by an empirical correction factor, $1 - 0.15 \exp(-(t_{\gamma}-30d)/25d)$, improves the estimate of the $^{56}$Ni mass to within 2\% or better. This correction depends on the time coverage of the  supernova light curve. For a real supernova, as below, we  derive this factor for its  actual observation sequence. For a high-significance measurement of a supernova light curve that includes late observations, we find that an accurate $^{56}$Ni mass can be derived from fitting this model to data only after the gamma-ray peak, although the parameter t$_{\gamma}$ then carries little information.

The parameter t$_{\gamma}$ is proportional to the square root of the mass of the overlying shell divided by its expansion speed, in the single-shell model. In a more realistic supernova with Ni at a range of depths, it is given by this ratio averaged over the $^{56}$Ni. Clearly, the information provided by this parameter can be used in a similar manner to that from t$_{0}$ derived by visible and infrared light curves \citep[e.g.,][]{2019MNRAS.483..628S}, but with fewer assumptions, and this additional information can be incorporated into such studies. A correlation of these two parameters could be established with high precision, perhaps further guiding SN~Ia modeling \citep{2020RNAAS...4..158S}.
We note that our t$_{\gamma}$ differs from the UVOIR derived t$_{0}$. The time when e$^{-1}$ of the line photons escape is later than when e$^{-1}$ of the gamma-ray energy escapes, because the escape of scattered photons enhances the total energy loss relative to 847 keV photon escape. For the typical formulation of t$_{0}$, the effective gamma-ray energy deposition opacity is  taken to be $\kappa=0.025 \, $cm$^{2} \, $g$^{-1}$ to convert to overlying mass. To the extent this simple model is appropriate in both energy ranges, we  expect the 
fitted parameters to be in the ratio $t_{\gamma}/t_{0}= ({0.069}/{0.025})^{1/2}$ = 1.66.

These fits are all done assuming complete coverage of the gamma-ray light curves to 400 days, i.e., a wide-field instrument with few observing direction constraints. In order to test the loss of information due to incomplete coverage, we discard at random a fraction  of the observation time for many simulated supernovae. If we retain at least one-third of the 400 days, the fitted parameters are not systematically affected, as long as there are some observations prior to the light curve maxima, and some observations at late times, after peak fluxes. The uncertainties on the parameters are increased with decreased total observing time approximately as $\Delta t^{-1/2}$, as expected. 

\begin{figure*}[ht!]
\plotone{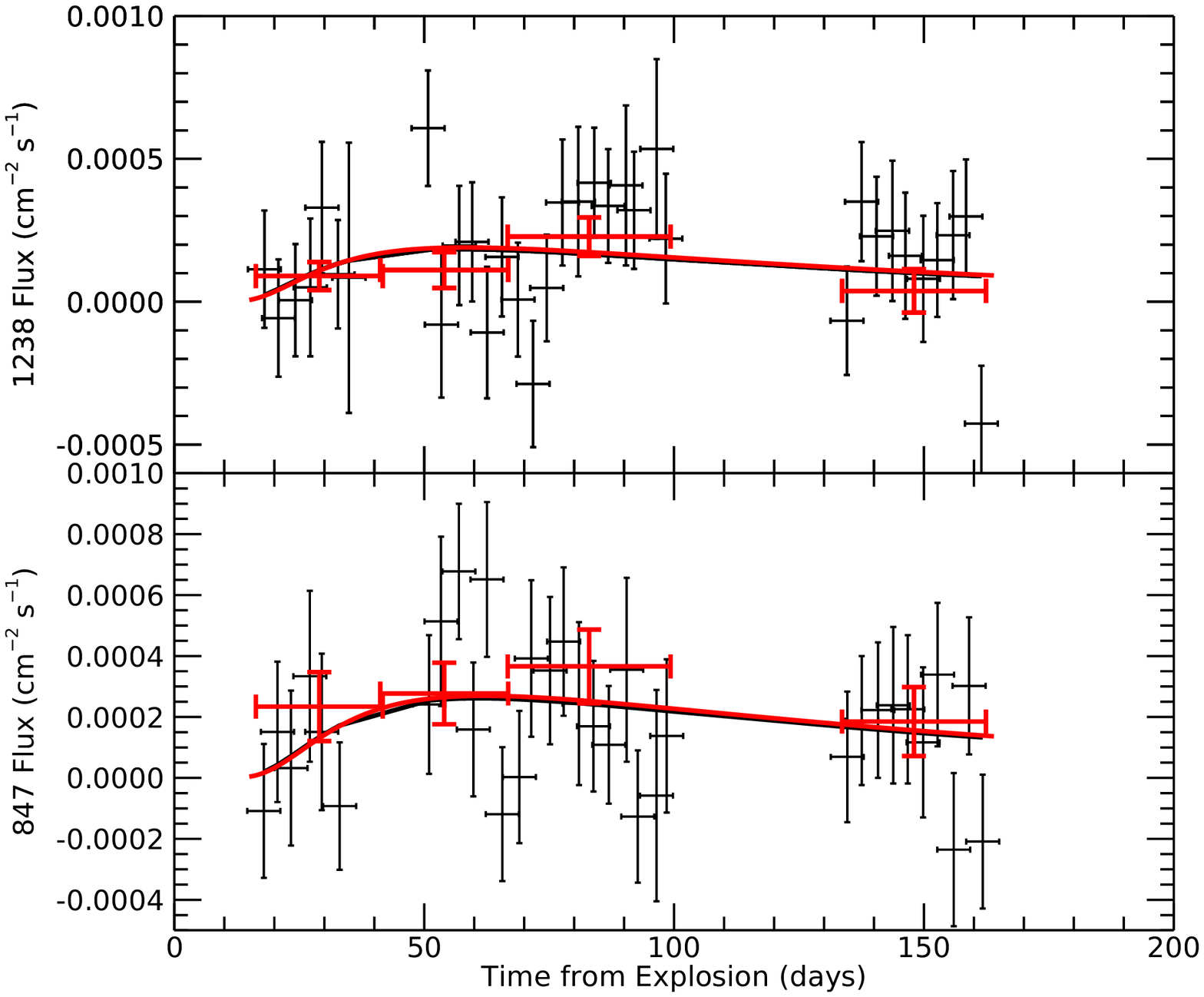}
\caption{Fits of the single shell, two parameter model to the two lines reported by  \citet[][red symbols]{2015A&A...574A..72D} and \citet[][black]{2015ApJ...812...62C}. The two data sets are fit independently, and the best-fit models are both plotted, almost entirely overlapping each other. \label{fig14j}}
\end{figure*}
\subsection{Supernova 2014J}
Only one SN~Ia has been detected in gamma-ray lines to date, SN 2014J 
\citep{2015A&A...574A..72D,2015ApJ...812...62C}. We fit this simple model to the reported 847~keV and 1238~keV band fluxes, treating them as line fluxes. These two studies used somewhat different background models and analysis methods, and report the results averaged over very different time bins. Our fit to the data of \citet{2015A&A...574A..72D} yields M($^{56}$Ni) $= 0.39 \pm 0.09$ M$_\odot$  (after correction) and t$_{\gamma}=32.4 \pm 8.8$ days. The fit is acceptable with $\chi^{2} = 4.94$ for six degrees of freedom. The fit to the fluxes reported by \citet{2015ApJ...812...62C} gives M($^{56}$Ni) $= 0.38 \pm 0.08$ M$_\odot$  and t$_{\gamma}=31.5 \pm 10.5$ days, with $\chi^{2} = 57.6$ for 64 degrees of freedom. Both data sets and fits are shown in Figure~\ref{fig14j}. Despite the fact that the data sets show some differences, the fits of this model to them are remarkably similar. Although we found above that models with t$_{\gamma}$ this small under-fit M($^{56}$Ni) by 14\%, we find that effect is reduced to 7\% for data sets that cover only the first 165 days, as these do, so we have applied that correction in the $^{56}$Ni masses stated. 

This M($^{56}$Ni) is somewhat small compared to the values quoted by these authors, 0.49$\pm$0.09 M$_\odot$  \citep{2015A&A...574A..72D} and 0.63$\pm$0.1 M$_\odot$  \citep{2015ApJ...812...62C}, which were mostly quantified by comparison to line light curves of previously published numerical models, and from other methods   \citep[e.g.,][]{2016A&A...588A..84D}. \citet{2015ApJ...812...62C} also compared the fluxes, including those of the continuum, to a grid of parameterized models, constraining  M($^{56}$Ni) and M$_{ej}$. These t$_{\gamma}$ values also at the  low end of all models we have fit, suggesting low total mass or large kinetic energy. The measured line fluxes, in particular, 847~keV, rise quickly and peak early, suggesting a relatively thin overlying mass and low total M($^{56}$Ni). This t$_{\gamma}$ corresponds to a 1 M$_{\odot}$ shell expanding at 12,000 km s$^{-1}$. This value for t$_{\gamma}$ is extremely low, considering that it is near the lowest t$_{0}$ values measured \citep{2014MNRAS.440.1498S}, but we expect t$_{\gamma}$ to be significantly larger (see above).
For the purposes of scaling future measurements, we note that based on the quoted error bars and time intervals, the INTEGRAL SPI sensitivity was approximately 5.0$\times 10^{-4}$ cm$^{-2}$~s$^{-1}$ for each of these two lines in 10$^{6}$ seconds, although \citet{2015A&A...574A..72D} show the 1238~keV sensitivity as somewhat better.

\section{Extracting Fundamental SN Ia Properties}

While the information we get from the single-shell model fits is limited, the fact that the fits are often not acceptable suggests that there is more information to be derived from the gamma-ray line light curves. Including measurements of the line profiles will also provide additional constraints. We seek to determine more fundamental parameters for a supernova, which would point toward specific progenitors and explosion mechanisms, for example, the  $^{56}$Ni mass, the total ejecta mass, ejecta kinetic energy, and the distribution of $^{56}$Ni within the ejecta. The escaping $^{56}$Ni and $^{56}$Co gamma-ray lines depend on the masses of those nuclei at a given location in the ejecta, and on the overlying total electron column density, integrated over the ejecta. In homologous expansion, all optical depths fall as time $t^{-2}$, and the escape of different energy lines are simply related by their Compton cross-sections. 

We take expansion velocity as the independent variable, as it is well constrained by visible spectra, and divide the supernova into a number of spherical shells, with expansion speeds proportional to radius, up to typically 20,000 km~s$^{-1}$. If we specify the mass density and initial $^{56}$Ni mass fraction of each zone, the gamma-ray line spectra are completely determined at all times. Here we seek to invert this procedure: given gamma-ray measurements of sufficient coverage and precision, how well can we determine the density and $^{56}$Ni abundance profiles, and therefore the fundamental properties of the ejecta? 

In the Appendix, we describe a fast method for determining gamma-ray line profiles and fluxes assuming only spherical symmetry and homologous expansion. Most SN~Ia are very nearly spherically symmetric \citep[e.g.,][]{2019NewAR..8701535S}. Homologous expansion sets in within seconds after the nuclear burning ends, but $^{56}$Ni decay could alter the velocity structure over 1--2 decay lifetimes \citep{2017ApJ...846...58H}. This might be detectable in only the nearest SN~Ia; typically, we are probing the ejecta properties after this time. 
While Monte Carlo methods for gamma-ray transport are widely used and accurate, we will typically need to iterate the ejecta parameters and evaluate the gamma-ray escape hundreds of times for a single data set in non-linear model fits, and thousands of times for Markov Chain Monte Carlo fitting. 

If we  specify the mass density throughout the ejecta, the escape of each gamma-ray line at each observer radial velocity at each time from every depth is then determined. We cast this as a matrix equation, where a row of a 2D matrix includes the transmission of the lines and decay of the isotopes at a given time, and the columns refer to different radial zones of the model. We multiply this matrix by the column vector of $^{56}$Ni masses  of the zones, and the resulting column vector contains the modeled observations versus time, a line flux in simplest form. We can also think of those elements as multiple line fluxes, or line spectral profiles, if fitting such data. The problem is linear in the $^{56}$Ni masses (or mass fractions, if the density is specified), and in principle this matrix can be inverted to give the $^{56}$Ni masses from the measured data. In practice, this matrix is ill-conditioned, so other linear techniques are used. 
\begin{figure}[ht!]
\plotone{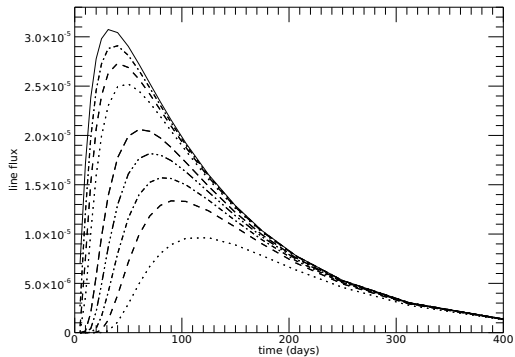}
\caption{The escaping gamma-ray line flux vs. time from ten spherical shells in homologous expansion, with mass density exponentially declining with expansion speed. The curves from top to bottom run from the outermost to innermost shells. The fluxes are for the the 847 keV line, in units of photons cm$^{-2}$~s$^{-1}$ per solar mass of $^{56}$Ni, at distance 20 Mpc.
 \label{bases}}
\end{figure}

Figure~\ref{bases} shows the columns of this matrix for a typical example, with exponential density profile and ten zones. This is for the 847 keV gamma-ray line flux, summed over its velocity profile. As basis functions for a fit, these are not nearly orthogonal, which represents a fundamental limit to extracting information from noisy data. Each of these curves is nearly the time dependence of transmission through a single overlying shell, as above, with different t$_{\gamma}$'s. They differ from exactly that because of the attenuation by the interior zones. The density distribution determines the shapes of the curves, the $^{56}$Ni abundances, their amplitudes. Because the shapes are slightly different for different cross-sections, the curves for different line energies are somewhat different, and multiple lines can help extract more information. Doppler information from line profiles can help break the linear dependence of the columns.

In analyzing real supernova gamma-ray data,  the density distribution should be determined from the data. The line fluxes are, of course, non-linear in the densities. Here we take multiple approaches. In the first, we use non-linear parametric models to fit both the mass density,  and $^{56}$Ni abundance distributions with velocity. In another, we use parametric models to fit the density, but non-parametric solutions of the $^{56}$Ni distribution. Applying physical constraints on the parameters is important in all these fits. 

\subsection{Parametric Models}

We use simulated data derived from the models as described above. First we extract line fluxes from spectral fits as described above, and for fitting, we sum the model line profiles over each line. Later we will keep the spectral lines to fit their profiles. Unless otherwise stated, we use a sensitivity to each line of 1.0$\times10^{-6}$ cm$^{-2}$ s$^{-1}$. The supernova properties we seek to determine are described by the functions $\rho(v)$ and X$_{56}(v)$. These two functions completely and simply determine all emergent gamma-ray line profiles at all times, once homologous expansion obtains. Examples of $\rho(v)$ used include exponentials, modified exponentials, power-laws, broken power-laws, Voigt profiles, Gaussians, and generalized Gaussians. Functions for X$_{56}(v)$ used include Gaussians, generalized Gaussians, and Voigt profiles. As one would expect, choosing functions that most closely match the distributions in the model simulated gives the best fits and most accurate results for a given data set here. However, these fitting functions are not fundamental, so we would choose the functions with the best fit. Here we will display only the most generally applicable combinations. Once the parameters of these functions are determined in a fit, we can sum the masses, the $^{56}$Ni masses, and the kinetic energies of the zones, to get the more meaningful constraints of total ejecta mass, total $^{56}$Ni mass, and total kinetic energy. 

For example, for a modified exponential density function and generalized Gaussian $^{56}$Ni mass fraction, we fit
\begin{equation}
\rho(v) = \rho_{0} \, e^{-(v/v_{0})^{a}}, \ \ \ \  X_{56}(v) = X_{0} \, e^{-(|v-v_{c}|/v_{g})^b}, \label{superg}
\end{equation}
calculating the line fluxes and comparing them to the simulated data. This fit has up to seven fitted parameters, $\rho_{0}$, $v_{0}$, $a$, X$_{0}$, v$_{c}$, v$_{g}$, and $b$, some of which must be constrained to be physically meaningful, i.e., $\rho_{0} > 0$, $0 \leq X_{0} \leq 1.0$, while others are constrained to be within broad reasonable ranges to aid in convergence of the fit. 

\begin{figure*}[ht!]
\plotone{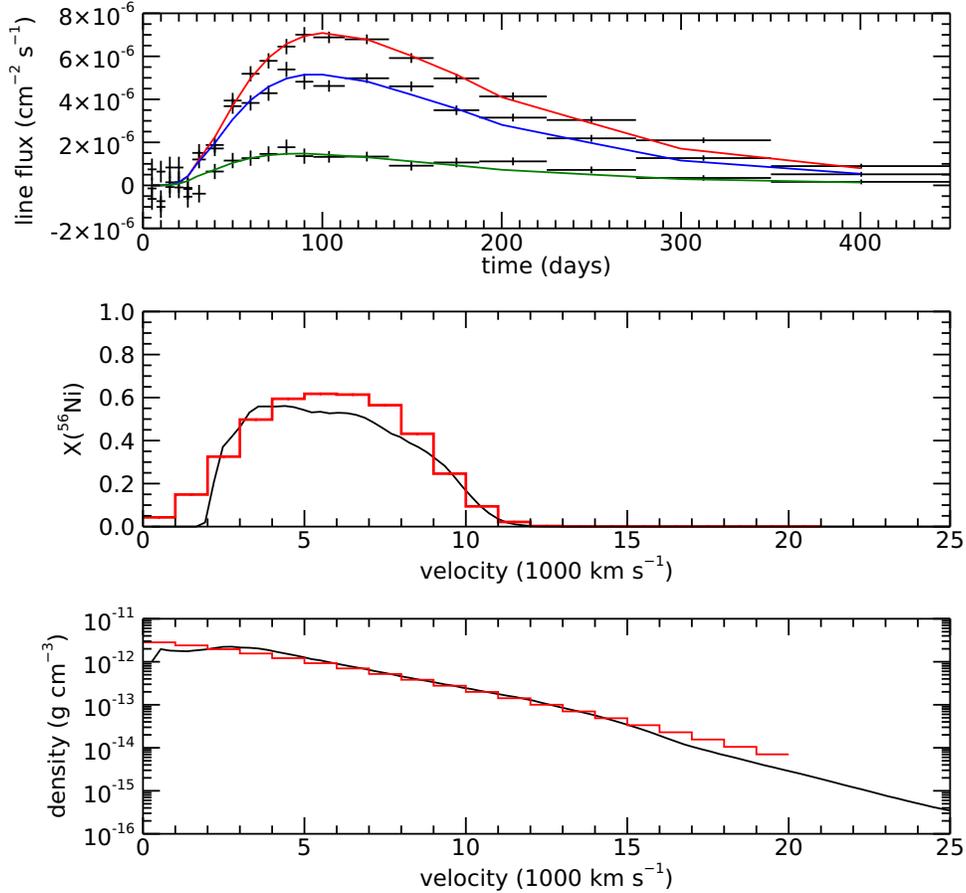}
\caption{A typical fit of the density and $^{56}$Ni mass fraction distributions of Equation~\ref{superg} to line fluxes derived from the model MWD at distance 20~Mpc. The upper panel shows the  847, 1238, and 2599 keV (top to bottom) line measurements, simulated from model MWD distributions, shown as black lines in the bottom two panels. 
Optimizing the parameters of our model to fit these data, we find the red histograms of the lower two panels, which give rise to the 
  fitted fluxes (red, blue, green, respectively) in the upper panel. The mass density in the lower panel is shown at 10$^{6}$ seconds after explosion.
 \label{example2}}
\end{figure*}

We show a typical fit of this form to data simulated from the model MWD at 20 Mpc in Figure~\ref{example2}. The fit to three line light curves (847, 1238, 2599 keV) is marginally acceptable with $\chi^{2}$=67.4 for 53 degrees of freedom. The density and $^{56}$Ni mass fraction distributions are reasonably, although not highly accurately, recovered. The derived ejecta mass, $^{56}$Ni mass, and kinetic energy, 1.93 M$_{\odot}$, 0.63 M$_{\odot}$, and 1.90 10$^{51}$~erg, respectively, are very close to the actual values (Table 1). We repeat this fit many times, changing only the statistical fluctuations of the data. The fits are statistically acceptable, but the parameters are strongly correlated, and their varying combinations can yield, e.g., a range of total masses. The $^{56}$Ni mass is constrained by the late data, but tends to be biased high by a few percent. The worst results, in terms of the supernova properties, occur when the outer layers are fit with high density and the $^{56}$Ni moved outward to appropriate optical depths to reasonably fit the data. Such a fit can yield total ejecta masses 30--50\% high in a fraction of the fits. In all of these fits, the line fluxes are relatively insensitive to the mass of the inner shells if there is no $^{56}$Ni there; those zones  scatter only a  small fraction of the emission from the receding sides of the outer zones. 

\begin{figure*}[ht!]
\plotone{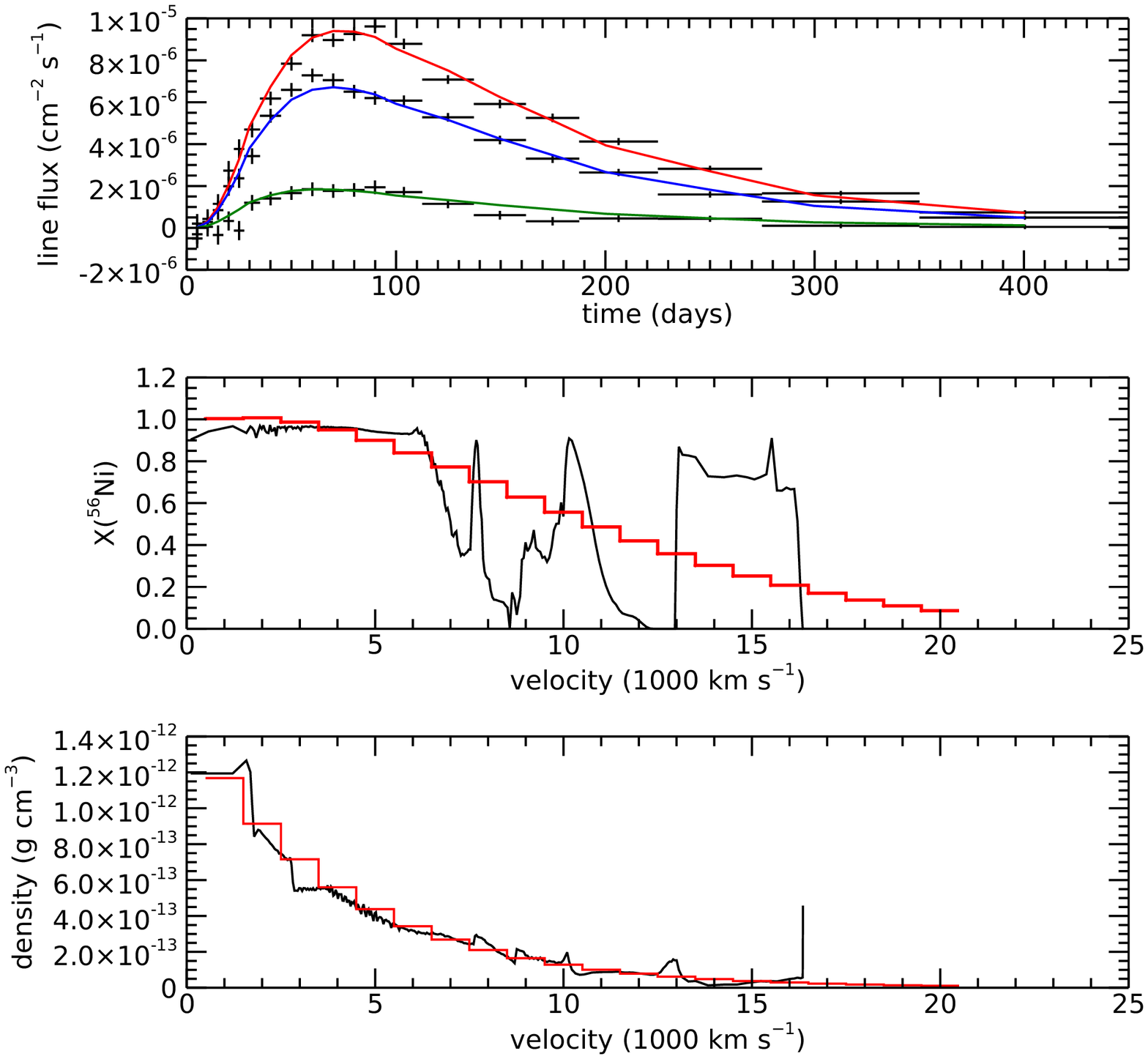}
\caption{A typical fit of the density and $^{56}$Ni mass fraction distributions of Equation~\ref{superg} to line fluxes derived from the model HeD8 at distance 20~Mpc. The upper panel shows the 847, 1238, and 2599 keV lines fluxes (top to bottom) and the fitted model fluxes (red, blue, green, respectively). The middle panel shows the $^{56}$Ni mass fraction of the fitted generalized Gaussian (red histogram), and the model HeD8 distribution (black line.) The bottom panel shows the fitted exponential density distribution (red histogram) and the HeD8 density (black line), at 10$^{6}$ seconds after explosion. Summing the fitted curves over the zones, the total mass, $^{56}$Ni mass, and kinetic energy are 0.98 M$_{\odot}$, 0.52 M$_{\odot}$, and $1.12\times 10^{51}$~erg, to be compared to the HeD8 values of Table 1.
 \label{example3}}
\end{figure*}

This situation can result in systematic uncertainties in the derived parameters beyond the expected statistical ones. We can improve the fits, at least avoiding extreme outliers, by incorporating prior information. A simple, effective constraint is to use the result of the single-shell fit of Section~\ref{sss} for the $^{56}$Ni mass, after correction. We multiply the likelihood by the fitted probability distribution for M($^{56}$Ni), which then regulates the fit parameters. Even after this improvement, we are limited by the choices of the velocity dependences of density and $^{56}$Ni mass. Another illustration of this is shown in Figure~\ref{example3}, where the data are derived from the SN~Ia model HeD8. The generalized Gaussian, or other simple shape, will not accurately describe such a distribution, but the more global properties are still reasonably recovered. 

We do not propagate the parameter uncertainties into those for the global SN~Ia properties here. Instead we repeat the process many times for a supernova model, varying only the fluctuations of the simulated data, to determine the uncertainties in the resulting derived quantities due to photon counting statistics. We use the identical setup of the non-linear fitting procedure and a single pair of functions. Given a smaller collection of real data sets, the fitting could be guided to provide, in some cases,  better convergence and fits to the data. Alternatively, Markov Chain Monte Carlo techniques can be used to explore the parameter space of each fit, but here we prefer fast analyses of many simulated data sets. 

\begin{figure*}[ht!]
\plotone{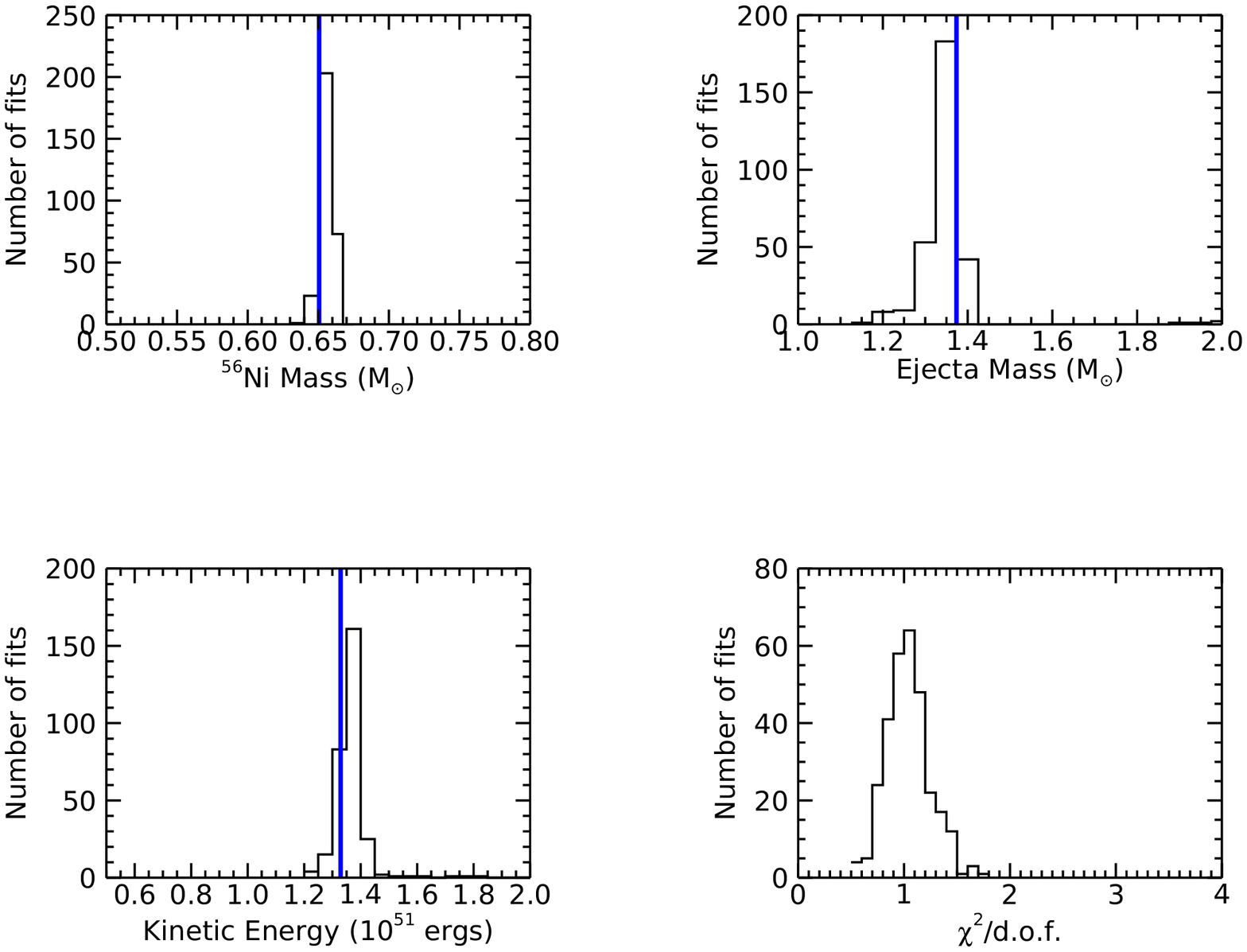}
\caption{Histograms of derived quantities from 300 fits of the functions of Equation~\ref{superg} to simulated line flux data derived from the model 7p0z22d20\_27 at 20 Mpc. Only the statistical fluctuations corresponding to the line sensitivities of 1.0$\times10^{-6}$ cm$^{-2}$~s$^{-1}$ were varied. The $^{56}$Ni masses are constrained by a prior derived from fits as described in Section~\ref{sss}. The blue vertical bars show the values of these quantities for the model 7p0z22d20\_27. 
 \label{d20hists}}
\end{figure*}

In Figure~\ref{d20hists} we show the results of repeating the fit using Equation~\ref{superg} to three line fluxes simulated from model 7p0z22d20\_27 three hundred times, changing only the Gaussian fluctuations in the ``measured'' spectra. Here we again use the $^{56}$Ni mass determined from the fit of Section~\ref{sss} to each realization of the data as a prior. We use the best-fit density and $^{56}$Ni mass fraction functions to sum the total ejecta mass, $^{56}$Ni mass, and kinetic energy for each instance. Histograms of these quantities are shown. We see that the ejecta mass and kinetic energy are well recovered, with some small biases due to the inaccuracy of these fit functions. The measured kinetic energy, in particular, is subject to a choice of the maximum velocity, where the sum is truncated. This is somewhat arbitrary here, as very fast, $^{56}$Ni-free, outer zones affect the measured gamma rays very little. For a real supernova, other spectroscopic measurements might provide the appropriate cutoff. The $\chi^{2}$ distribution has a reasonable shape, and is skewed to slightly higher values than expected from statistical fluctuations alone. There are a few outliers at high ejecta mass and kinetic energy, which are typically also those with the worst $\chi^{2}$, and at least in some cases can be improved by varying the initial parameter guesses or convergence parameters in the non-linear fit. Our goal here is not to perfect the fit to this supernova model, but to estimate the precision of this method of determining the physical properties of the supernovae. 

Summing  the histograms around their peaks give one-sigma uncertainties for the ejecta mass and kinetic energy of 0.05 M$_{\odot}$ and 0.07$\times10^{51}$ ergs, respectively. Repeating this exercise for the other supernova models gives similar results. Some other information can be derived from such fits. For example, the outer velocity at which X($^{56}$Ni) falls to 0.5 is recovered well, to within 2,000--3,000 km s$^{-1}$, in this case. This depends on the appropriateness of the fit functions; this is  the case for for 7p0z22d07\_27 and MWD, but not the supernova model HeD8. 

Of course, the precision with which supernova properties can be determined depends on the signal-to-noise of the measurements, i.e., on the distance and instrument sensitivity. We test this dependence by changing either one. For example, at larger distance, the systematic effects of the imperfect fit model are reduced, but the $^{56}$Ni mass determined in the first step is less precise, and therefore the prior constraint is less effective. This latter effect dominates and the result is that the precision of the determinations of all of $^{56}$Ni mass, total ejecta mass, and kinetic energy vary with the signal-to-noise of the line flux measurements. We would expect that the uncertainty in the total ejected mass, for example, goes as
\begin{equation}
\sigma M_{ej} = 0.05 \, (\frac{D}{20 \, Mpc})^{2} \, (\frac{F_{3\sigma}}{10^{-6} \, cm^{-2} \, s^{-1}}),
\end{equation}
where F$_{3\sigma}$ is the sensitivity to each of the three lines used here. We find a somewhat shallower variation. For lower S/N data, more fits fail to converge, and the results are improved by fixing or tightly constraining, some parameters. For example, at 60 Mpc, we fix the parameters $a$ and $b$ at typical values, 1.0 and 3.2, respectively. For  a rare, much nearer supernova, the fits here are statistically quite poor, and systematic effects dominate the precision of the measurements. The success of this method will then depend on finding density and radioactivity distributions that fit the data well. 

\subsection{Nonparametric  $^{56}$Ni distributions}

While models often indicate, and some observations are consistent with, roughly exponential density decline with velocity,  the  $^{56}$Ni radial distributions are comparatively unconstrained. Because the line fluxes are linear in the zonal $^{56}$Ni masses, we can in principle derive their profiles from the data using linear inversion techniques. We still must fit the mass density using techniques of the previous section. We use the variable projection method \citep[e.g.,][]{2003InvPr..19R...1G}  to solve this problem, where the linear parameters are solved for in each iteration of the nonlinear parameters, which are  found using the Levenberg-Marquardt method in nonlinear least squares. In order to avoid introducing too many additional, albeit linear, parameters, we use coarse radial zoning of the ejecta. For the solution of the linear problem, we use either singular value decomposition (SVD; \citet{2002nrca.book.....P}) or Bounded Variable Least Squares (BVLS; \citet{80008364165}). In either case, the mass of $^{56}$Ni of each zone is a parameter bounded by zero and the total zone mass. 

\begin{figure*}[ht!]
\plotone{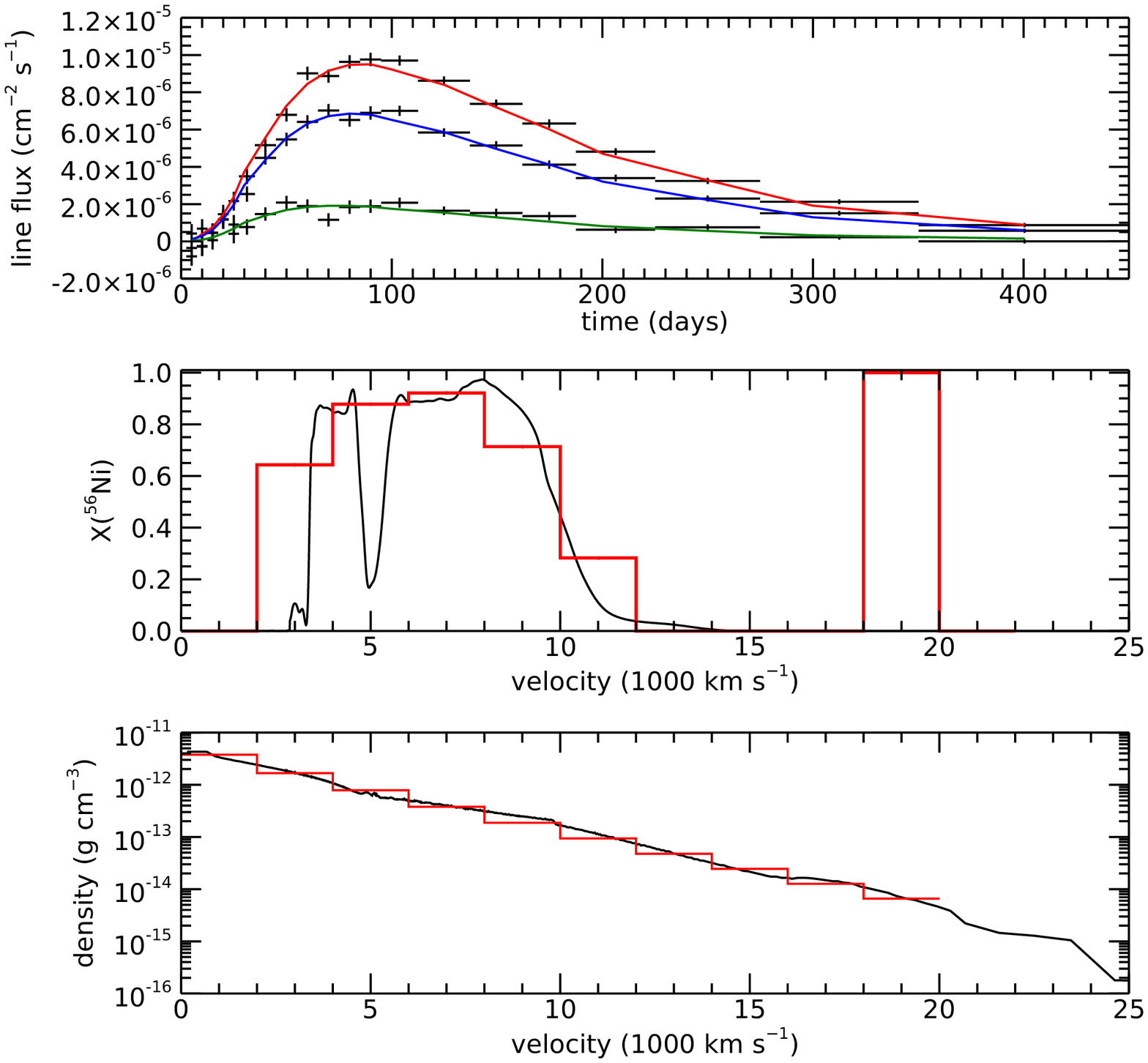}
\caption{A  fit of the density, with a modified exponential, and $^{56}$Ni abundance, with ten mass fractions, distributions to line fluxes derived from the model 7p0z22d20\_27 at distance 20~Mpc. The upper panel shows the 847, 1238, and 2599 keV lines fluxes (top to bottom) and the fitted fluxes (red, blue, green, respectively). The middle panel shows the $^{56}$Ni mass fraction found using SVD (red histogram), and the model 7p0z22d20\_27 distribution used to generate the data (black line.) The bottom panel shows the fitted modified exponential density distribution (red histogram) and the 7p0z22d20\_27 density (black line), at 10$^{6}$ seconds after explosion. 
 \label{d20svd}}
\end{figure*}

\begin{figure*}[ht!]
\plotone{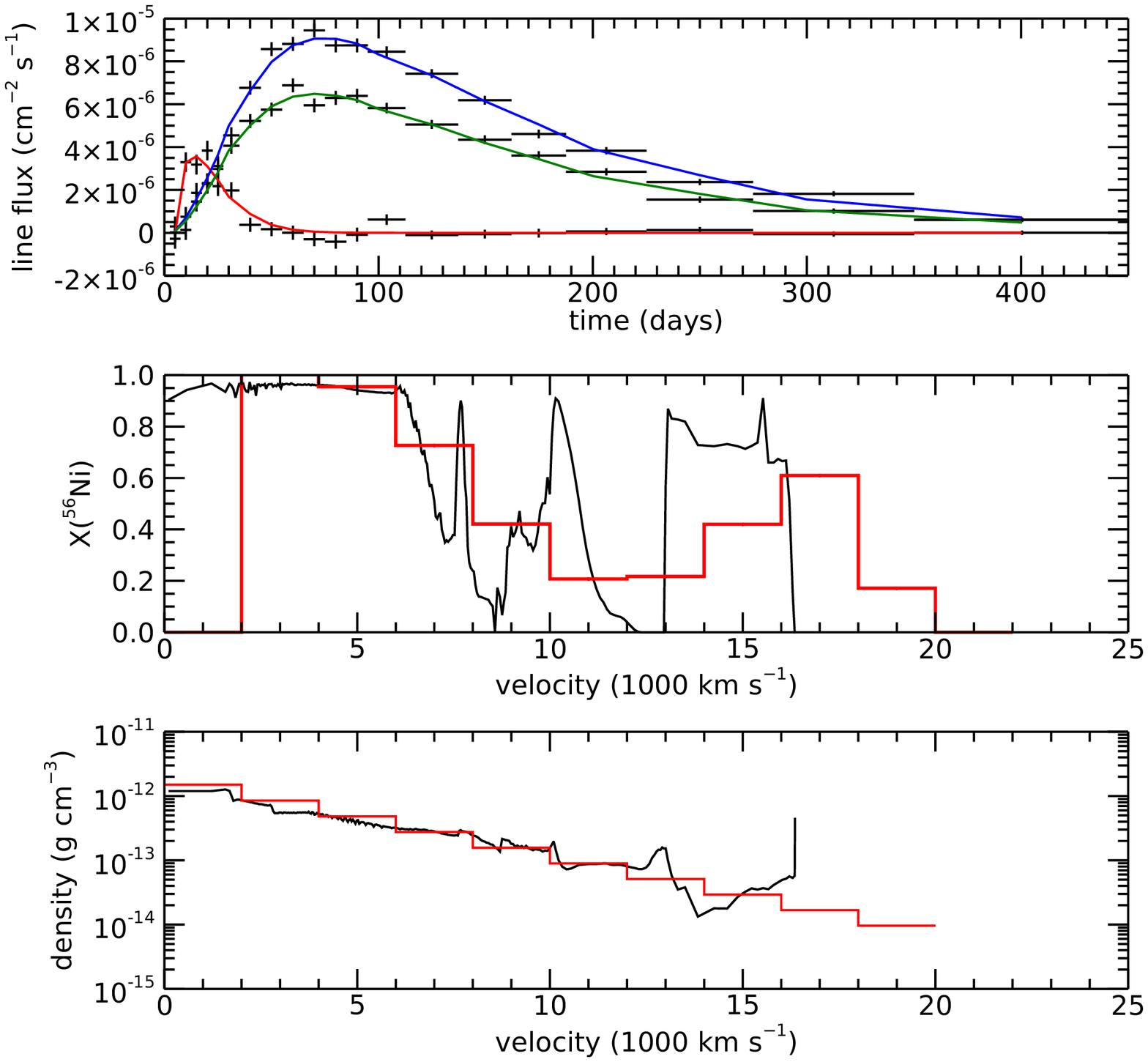}
\caption{Similar to Figure \ref{d20svd} but for data simulated from the supernova model HeD8. Four line fluxes were fit, but 847, 1238, and 812 keV lines are shown, top to bottom.  
 \label{hed8svd}}
\end{figure*}

To illustrate this method, we show in Figure~\ref{d20svd} a typical fit to line fluxes simulated from 7p0z22d20\_27 at 20 Mpc of a model with a modified exponential density distribution and   $^{56}$Ni mass fractions for each of ten zones, constrained between zero and one, found using SVD. The details of the supernova model used to generate the data are overplotted. With these additional parameters, we typically get good fits, for any signal-to-noise. Summing over the zones, the $^{56}$Ni mass, ejecta mass, and kinetic energy are very close to the 7p0z22d20\_27 values. The large $^{56}$Ni abundance in the outermost zone is a common occurrence in these fits. We find that if we add the data for the $^{56}$Ni 812 keV line, even though the fluxes are quite low, the abundance in the outer zone is greatly constrained. We also show in Figure~\ref{hed8svd} the same model fit to data simulated from the supernova model HeD8. At this signal-to-noise, we do not reliably recover all the details of the input $^{56}$Ni distribution, as the $^{56}$Ni can be moved among neighboring zones with little effect on the overall goodness of fit. 

\subsection{SN 2014J}   
For illustration, we apply these two types of fits to the 847 and 1238 keV line fluxes of  \citet{2015ApJ...812...62C}, assuming a distance of 3.5 Mpc. 
 Because of the relatively low signal-to-noise of the measurements, we minimize the number of parameters by using the simplest possible functions, for example, a simple exponential density distribution (i.e., the parameter a=1 in Eq.~\ref{superg}). We try fits with $\rho(v)$ represented by exponential, power-law, Voigt, linear, Gaussian, and broken power-law functions. For X($^{56}$Ni) we use a simple Gaussian (i.e., Equation~\ref{superg} with b=2), and eight separate abundances, determined by SVD or BVLS. We do not apply a prior to the $^{56}$Ni mass, as it turns out to have no effect here. 
 
None of these combinations stands out as a significantly better fit than the others. These models improve the fit $\chi^{2}$ by only about unity relative to the simpler model described above, hardly justifying the additional parameters. Because these are more physically realistic, we describe a few of the results. For exponential density functions, the best-fit scale length is 3,000 to 4,500 km s$^{-1}$, but is  poorly constrained. This parameter is highly anti-correlated with the $\rho_{0}$ parameter, so the total mass, for example, is much better constrained. Best-fit Gaussian X($^{56}$Ni) are centered at 7,000--9,000 km s$^{-1}$, with FWHM approximately 7,000 km s$^{-1}$, depending on the density function used. SVD-determined X($^{56}$Ni), with singular value cutoff such that the rank of the modified matrix is four, is near unity for 5,000 -- 13,000 km s${-1}$, and very small elsewhere. The BVLS solutions show similar, but slightly narrower, high values centered around 9,000 km s$^{-1}$. 

Best fits to all of these model combinations give  M($^{56}$Ni) values near 0.4 M$_\odot$. The most discrepant value is 0.46 M$_\odot$  for a fit with a power-law density distribution and SVD solution to the X($^{56}$Ni) array. Total ejecta masses vary somewhat more, from a low value of 0.73 M$_\odot$  (Gaussians for both functions) to 1.10 M$_\odot$  (power-law density with Gaussian X($^{56}$Ni)). Most combinations cluster around 0.85 M$_\odot$  at best-fit. Kinetic energies vary substantially, from 0.7 to 2.0$\times 10^{51}$~erg among the fit combinations. We have little justification for choosing among these fit functions, but to illustrate the precision of these determinations, we look closer at the model using exponential $\rho(v)$ and SVD determination of the X($^{56}$Ni) values. We evaluate $\chi^{2}$ on a fine grid of the density parameters surrounding the best-fit values, marginalizing over the $^{56}$Ni distribution by finding the best-fit array for each grid point. We then calculate M($^{56}$Ni), M$_{ej}$, and kinetic energy at each point. We find M($^{56}$Ni) = 0.39$^{+0.11}_{-0.08}$ M$_\odot$, M$_{ej}$ = 0.84$^{+0.15}_{-0.28}$ M$_\odot$, and KE = 1.04$^{+0.20}_{-0.59}$ 10$^{51}$~erg, where all uncertainties are 1$\sigma$ values. For one comparison, we employ 
a Markov Chain Monte Carlo (MCMC) search of the parameter space of an exponential density distribution and Gaussian X($^{56}$Ni). Fixing the Gaussian amplitude at the best-fit value 0.90 and following ten chains of 2,000 steps each in the remaining four-dimensional space, we find uncertainties very close to those quoted just above. 

All our analyses of the $^{56}$Ni lines alone point to a $^{56}$Ni mass near 0.40 M$_\odot$, and an ejecta mass below 1 M$_\odot$  for SN 2014J. However, we cannot exclude a Chandrasekhar mass ejecta at high confidence. Given the precision of the INTEGRAL-SPI measurements, we cannot choose among many fitting functions, constrain well the parameters of any one, or determine $^{56}$Ni distribution, but it is clear that an instrument with 100--500 times better sensitivity could provide detailed constraints on many SN~Ia. 

\section{ Fitting Line Profiles} \label{sec:profiles}

The above analyses apply to any gamma-ray detector that can extract line fluxes, i.e., with energy resolution of order $\Delta E/E = 10\%$ or better. Instruments with better energy resolution and sufficient sensitivity to measure line profiles can add velocity information to further constrain supernova parameters. This has long been discussed, but a method has not been demonstrated. As shown in the Appendix, from mass density and $^{56}$Ni abundance distributions, and assuming only spherical symmetry, we get the full gamma-ray line intensity profiles of all lines at all times, which can be compared directly to measured line spectra. 

Here we define the instrument by its {\it narrow} line sensitivity and its energy resolution. Instruments with good energy resolution have improved narrow line sensitivity for a given continuum background, because they resolve the background and include less of it under the line. However, for intrinsically broad lines, such an instrument's line sensitivity degrades, as more background counts appear under the line. A high energy resolution detector therefore has worse sensitivity to a broad (e.g., SN~Ia) line than a poorer resolution detector with the same narrow line sensitivity. We use the same calculated Monte Carlo spectra of physical supernova models as above. We convolve them with a Gaussian instrument energy resolution, rebin them to a specified number of bins per detector resolution element, and add noise fluctuations appropriate to the quoted sensitivity and binning. Before these steps, we subtract a flat Compton continuum below each line center, because we do not calculate the continuum in closed form like the lines. There is information on the optical depth traversed by the photons in this continuum, but we do not use it here. One would analyze much broader energy ranges to take full advantage of this information \citep{2014ApJ...786..141T}.
\begin{figure*}[ht!]
\plotone{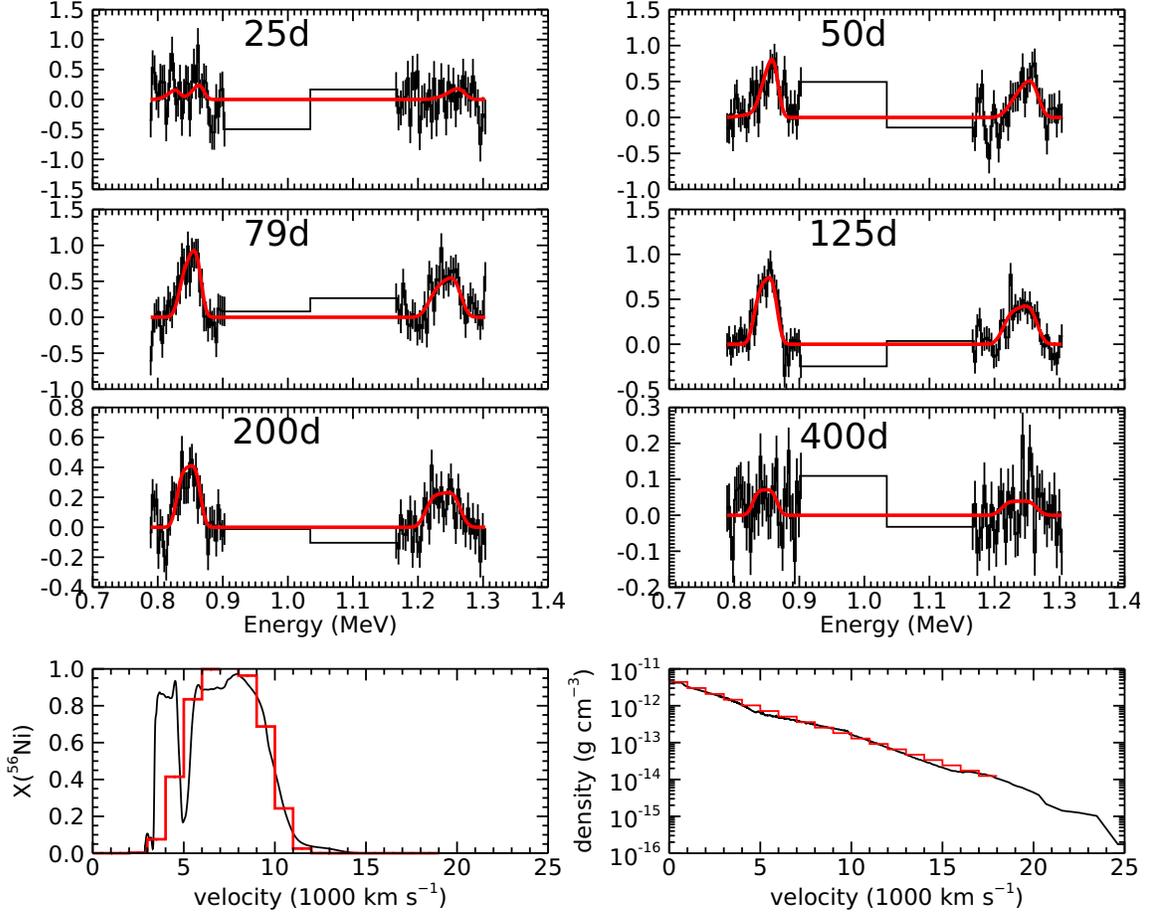}
\caption{A typical fit of the density and $^{56}$Ni mass fraction distributions of Equation~\ref{superg} to line profiles derived from the model 7p0z22d20\_27 at distance 20~Mpc. This is for an instrument with energy resolution $\Delta$E/E = 1\% and 3$\sigma$ sensitivity to narrow lines 1.0$\times 10^{-6}$ cm$^{-2}$~s$^{-1}$ in 10$^{6}$ s. Here we fit the 847 keV and 1238 keV line regions. The 812 keV line must be included to fit the early 847 keV profiles. All twenty observation periods are fit, but only a few are shown.  The fitted supernova parameters and resulting line intensity profiles are shown in red. In the bottom panels, the  black curves show the model 7p0z22d20\_27 distributions from which the simulated data were derived.
 \label{spec1}}
\end{figure*}

\subsection{Parametric Models}
The supernova density and $^{56}$Ni abundance distributions in velocity are defined as above, for example, with density as a modified exponential, and abundance as a generalized Gaussian. Then we fit the spectra around the chosen line energies for all (up to twenty) observation periods, adjusting the supernova parameters to optimize the fit. To illustrate this, in Figure~\ref{spec1} we show a fit to the 847 keV and 1238 keV regions for data simulated from model 7p0z22d20\_27 at 20 Mpc with an instrument with $\Delta$E/E = 1\% and narrow line sensitivity 1.0$\times 10^{-6}$ cm$^{-2}$~s$^{-1}$. All energies and times are fit simultaneously. For any early observations, the $^{56}$Ni 812 keV line must  be modeled along with the 847 keV line. We find that for energy resolution roughly 4\% or better, we recover the input $^{56}$Ni mass and total mass without using any prior from line flux fits. Using the available velocity information distinguishes the solution from fits with more mass and the $^{56}$Ni in the outermost layers, even for modest energy resolution.
\begin{figure*}[ht!]
\plotone{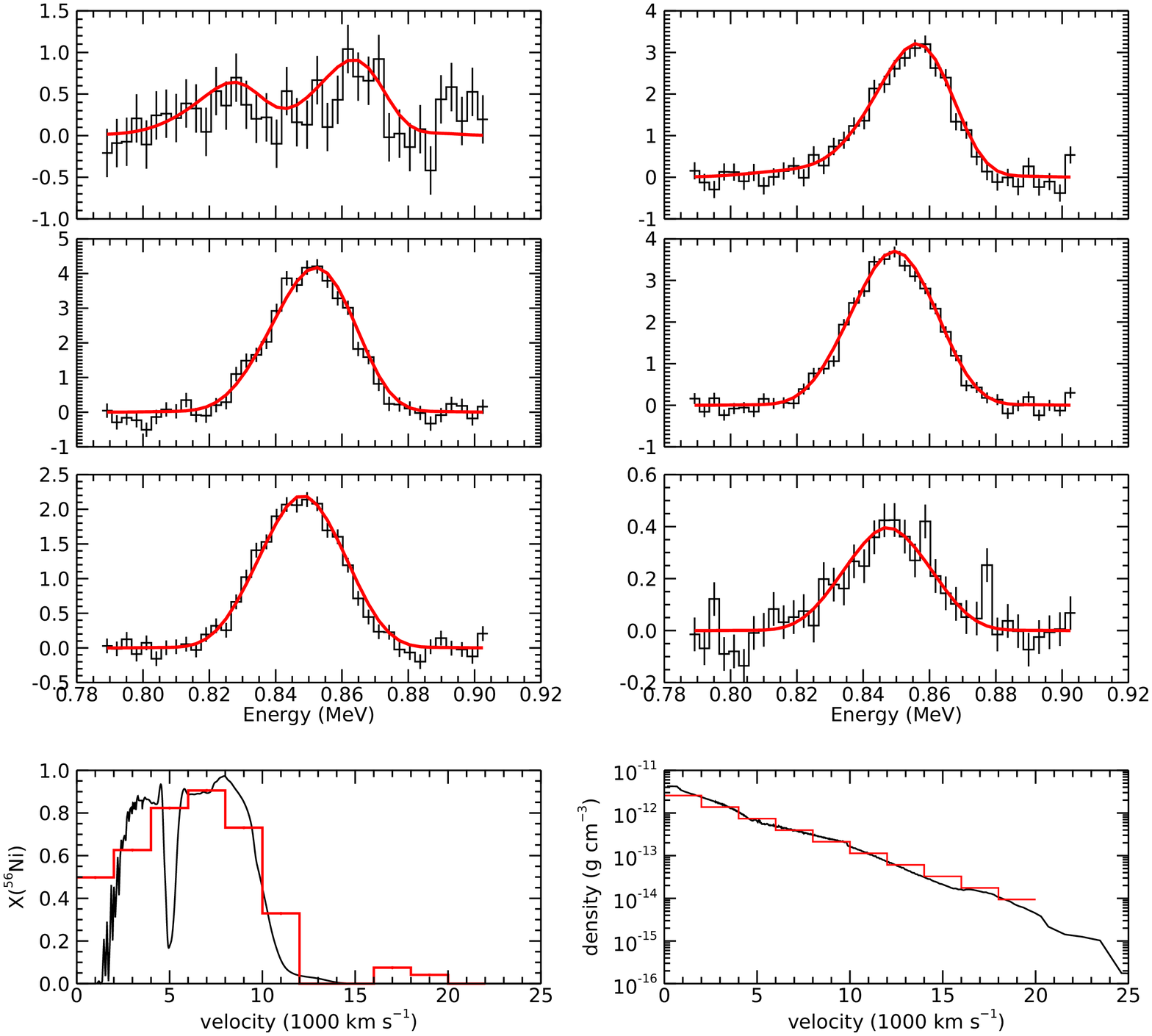}
\caption{A fit of the density modeled as a modified exponential and $^{56}$Ni mass fraction as ten separate values to the 847 keV region derived from the model 7p0z22d07\_27 at distance 10~Mpc. This is for an instrument with energy resolution $\Delta$E/E = 1\% and sensitivity to narrow lines 1.0$\times 10^{-6}$ cm$^{-2}$~s$^{-1}$ in $10^{6}$ s.   Twenty observation periods are fit, but only a few are shown.  The fitted supernova parameters and resulting line intensity profiles are shown in red. In the bottom panels, the  black curves show the model 7p0z22d07\_27 distributions from which the simulated data were derived. The spectra are in units of 1.0$\times 10^{-6}$ photons cm$^{-2}$ s$^{-1}$ per bin.
 \label{spec2}}
\end{figure*}

\subsection{Non-parametric Models}
We can also extract nonparametric $^{56}$Ni abundance distributions from line profile data, using the same methods as above for fluxes alone. In an optimistic example, we show in Figure~\ref{spec2} a typical fit to the 800--900 keV spectrum derived from the model 7p0z22d07\_27 placed at 10 Mpc. The instrument is defined by narrow line sensitivity 1.0$\times 10^{-6}$ cm$^{-2}$~s$^{-1}$ (3$\sigma$) and energy resolution $\Delta$E/E = 1\% FWHM. We fit the three parameters of a modified exponential along with ten $^{56}$Ni abundances, determined from SVD within the variable projection method mentioned above. With this signal-to-noise, the supernova model distributions are accurately recovered, as are their integrals, ejecta mass, $^{56}$Ni mass, and kinetic energies. Again, fitting line profile data relieves the need for prior constraints, and nearly eliminates poor fits and those with outlying values of supernova properties. 
\begin{figure*}[ht!]
\plotone{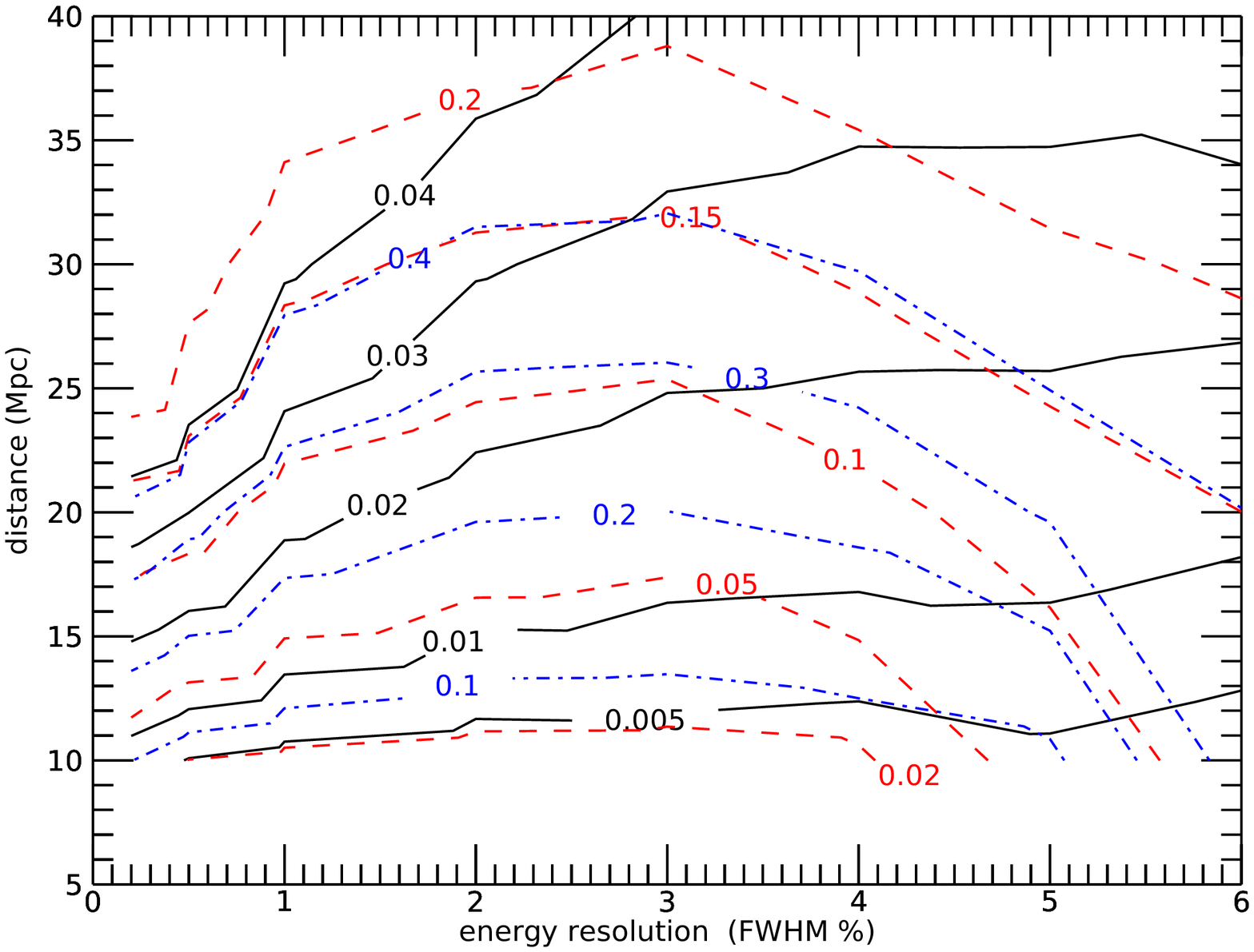}
\caption{Contours of the uncertainties on the SN~Ia properties $^{56}$Ni mass (black), ejecta mass (red), and kinetic energy (blue) determined from fits to the model d07, as a function of distance and instrument energy resolution, for an instrument with narrow line sensitivity 1.0$\times 10^{-6}$ cm$^{-2}$~s$^{-1}$ in 10$^{6}$ seconds. At each point in this plane, we do 100 fits varying only the noise in the spectra. Contour labels show the 1$\sigma$ uncertainties in units of M$_{\odot}$ for the masses and 10$^{51}$ ergs for kinetic energy. Here we use a modified exponential for $\rho(v)$ and SVD to determine X($^{56}$Ni).
 \label{grid1}}
\end{figure*}
We again measure the sensitivity of our determination of $^{56}$Ni mass, ejecta mass, and kinetic energy by repeating a fit like that shown in Figure~\ref{spec2} many times, varying only the statistical fluctuations in the simulated data. We repeat this for our four supernova models, at several distances, and for several instrument energy resolutions. 

We find, for all SN~Ia models, that a better energy resolution instrument of a given narrow line sensitivity, though it suffers in broad line sensitivity, effectively recovers that loss in sensitivity for good signal-to-noise measurements for this analysis. The uncertainties on the supernova bulk properties are nearly independent of energy resolution for a given narrow line sensitivity, for high signal-to-noise. The added velocity information overcomes the additional background fluctuations included under the broad lines. At poorer signal-to-noise, the precision follows closely the broad line sensitivity for resolved lines. 

To illustrate this, we calculate the widths of the distributions of those supernova properties for a grid of distances (or, equivalently, sensitivities) and energy resolutions, fitting one hundred events with varying statistical fluctuations in the spectra, at each grid point. No priors are included in the fits. 
The results are shown in Figure~\ref{grid1} for data derived from the model 7p0z22d07\_27. There we show contours of the 1$\sigma$ uncertainties in $^{56}$Ni mass, ejecta mass, and kinetic energy as a function of density and energy resolution. The uncertainties are quoted as one-half the range in the parameter that encloses 32\% of the fits above and below the most probable value. For energy resolution 3\% and below, the most probable values are well within 1$\sigma$ of the underlying model values, except for kinetic energy, which tends to be biased high by roughly 1$\sigma$. Above 3\% energy resolution, the accuracy also suffers, but prior information could be applied as above to improve this. Again, for a different instrument sensitivity, F$_{lim}$, in units of $10^{-6}$ cm$^{-2}$~s$^{-1}$, one can get the distance at which a given contour is achieved as $D / F_{lim}^{1/2}$, where D is the distance plotted in Figure~\ref{grid1}. 
Fits to data from other supernova models give similar results. Systematic offsets are more frequent and larger for models whose distributions are less well approximated by the fit functions (e.g., whose density is not exponential in velocity.)

\subsection{Pointed Instruments}
In the preceding sections, we assume good coverage of the SN~Ia gamma-ray light curves. This is essential for using the total line fluxes to probe supernova properties, and is a common capability of many modern instrument concepts \citep{2021ExA....51.1225D,2019BAAS...51g.123M}. However, it might be feasible to improve sensitivity, as in most of astronomy, by employing large-area collectors to focus gamma rays on small, low background, detectors. While mirrors remain problematic at the energies we consider here, lenses can in principle be used \citep{1992ExA.....2..259L,2009ExA....23..121K}.  A large space observatory with the capabilities we require would likely have many objectives, or at least many supernovae to observe, a small field-of-view, and limited target pointings. Therefore, most SN~Ia would be observed a very small number of times. While this is insufficient to fit line flux light curves,  we investigate how much information can be derived from the line profile(s) in just a few measurements.

Even for excellent narrow line sensitivity of 3$\times 10^{-7}$ cm$^{-2}$~s$^{-1}$ (3$\sigma$), a single 10$^{6}$~s observation of a line profile from a supernova at 20 Mpc recovers the input supernova properties rather poorly in our analysis. This is true regardless of the time of the observation. However, as few as two such observations can improve the situation greatly. For the aforementioned quantities, two 10$^{6}$~s observations can be fit to recover accurately the $^{56}$Ni and total ejecta masses. Simulating all four models we use, and repeating each fit 100 times, we accurately recover M($^{56}$Ni) with 0.02 M$_\odot$  or better precision, and M$_{ej}$ to better than 0.1 M$_\odot$. Kinetic energy is less well recovered, to 0.3--0.4 10$^{51}$~erg, due to the strong correlations of, especially, density falloff with other parameters. These are quoted for Gaussian X($^{56}$Ni), but we can also derive nonparametric $^{56}$Ni profiles, recovering the main features of the input model $^{56}$Ni distribution, if not  its detailed structure.

The timing of the two observations is important. Optimally, both occur while the signal is large, but they should be separated such that the line is still blue-shifted in one and nearly centered at the rest energy in the other. For the models we simulate, one observation around 75 days and another around 150 days works well. The precisions quoted above result for energy resolution from 1--3\%. Above and below those, the precision degrades for a given narrow line sensitivity, and above, the accuracy suffers as well. Note that there are no gamma-ray derived prior constraints available in this scenario. 

\subsection{COSI}
\begin{figure*}[ht!]
\plotone{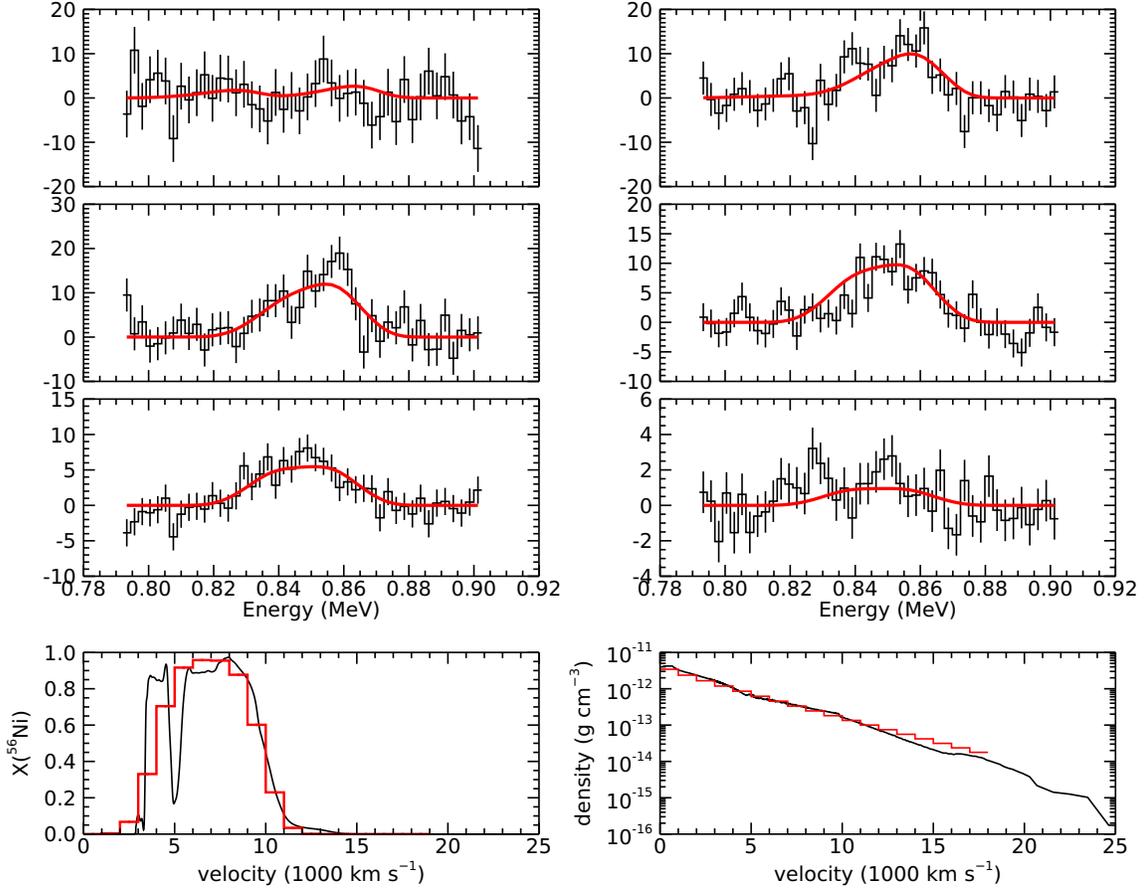}
\caption{A fit  of the density and $^{56}$Ni mass fraction distributions of Equation~\ref{superg} to line profiles derived from the model 7p0z22d20\_27 at distance at distance 5~Mpc. The three brightest lines are fit, but only the 847 keV region is shown for clarity. Our simple instrument parameters correspond to the line sensitivities and energy resolutions of the COSI requirements. COSI will observe the entire sky daily, so we fit twenty observation periods over 400 days, but only a few are shown.  The fitted supernova parameters and resulting line intensity profiles are shown in red. In the bottom panels, the  black curves show the model 7p0z22d07\_27 distributions from which the simulated data were derived. The spectra are in units of 1.0$\times 10^{-6}$ photons cm$^{-2}$ s$^{-1}$ per bin.
 \label{cosi}}
\end{figure*}
For a final example, we consider the {\em Compton Spectrometer and Imager}, COSI 
\citep{2021arXiv210910403T}. COSI is the only MeV gamma-ray instrument currently approved for the near future. It is a small, compact instrument, designed for study of long-lived galactic radioactivity, electron-positron annihilation, and polarization of bright sources, but it will have significant capability for studying nearby SN~Ia. We show in Figure~\ref{cosi} an example of a fit to data simulated from model 7p0z22d20\_27 at 5 Mpc with an instrument with COSI's line sensitivity and excellent energy resolution. While SN~Ia are rare at this distance, there is a good chance of one or more during its mission lifetime. The precision with which it would measure SN~Ia properties at this distance is nearly as good as our nominal examples at 20 Mpc, where we assumed line sensitivities 20--30 times better.

\section{Summary}

We illustrate a method for fast calculation of emergent gamma-ray line profiles from spherically symmetric expanding ejecta embedded with radioactive $^{56}$Ni, with no additional assumptions. We simulate gamma-ray measurements for specified published supernova models, distances, and instrument characteristics. We fit those data with lines calculated from parameterized density profiles, and either parameterized or non-parametric $^{56}$Ni abundance profiles. We integrate over density and $^{56}$Ni abundance to quantify $^{56}$Ni masses, ejecta masses, and kinetic energies inferred from the fits. We apply this to fit the full line profiles, or sum over the lines and fit fluxes alone versus time. We find:
\begin{itemize}
\item With sufficient signal-to-noise the velocity profiles of density and $^{56}$Ni abundance of the models used to simulate the data are well recovered from line profile and even from line flux measurements. The accuracy is limited by the appropriateness of the distributions fit, for example the $\rho(v)$ function.
\item Ejected $^{56}$Ni mass can be determined to 0.02 M$_\odot$  or better for our fiducial instrument sensitivity for a few SNIa per year, and better than current measurement precision for many. 
\item Total ejecta mass can be determined to better than 0.1 M$_\odot$  for those same supernovae. 
\item Ejecta kinetic energy is not as well determined, with a precision near 0.25$\times10^{51}$~erg.
\item Radioactive $^{56}$Ni abundance profiles vs. velocity can be accurately recovered for achievable signal-to-noise. 
\item For gamma-ray instruments with energy resolution 3--4\% or better, line profile fits are more robust than line flux fits, although the latter can be significantly improved with a prior constraint on $^{56}$Ni mass from a simple model fit. The measurement precision for SN Ia properties approximately follows the narrow line sensitivity.  
\item Using line flux measurements at 847 and 1238 keV from INTEGRAL/SPI \citep{2015ApJ...812...62C,2015A&A...574A..72D}, our analysis favors low $^{56}$Ni mass and low total ejecta mass relative to other estimates. We find from multiple analyses that M$(^{56}Ni)$=0.40$\pm0.10 \ M_{\odot}$ and M$_{ej}$=0.85$\pm0.20 \ M_{\odot}$, at 68\% confidence. 
\end{itemize}

While here we consider the information we can derive from gamma-ray line measurements alone, these will be used in concert with other approaches and information. Measuring the $^{56}$Ni mass to high precision can aid in the understanding of  extinction and reddening by dust, for example. Adding ejected mass measurements can inform challenging radiative transfer calculations in the UVOIR bands. Properties of supernovae derived as described here can serve as starting points for more detailed modeling, to be simulated a smaller number of times with Monte Carlo transport calculations, which can be tested against the full observed gamma-ray spectra, including scattered continuum. 

\begin{acknowledgments} The author is grateful to Peter H\"oflich for providing supernova calculations and for helpful discussions, and to Peter Milne and Lih-Sin The for careful review of the manuscript. The author gratefully acknowledges decades of valued discussions of this and related subjects with Donald D. Clayton. 
\end{acknowledgments}

%
%

%



\appendix

\section{Gamma-ray Line Escape}

Assuming spherical symmetry and homologous expansion, the transfer of gamma-ray lines in a supernova is simple in comparison to other supernova line calculations. Compton scattering  removes  photons from the line like a true absorption process, unless the scattering angle is so small as to have no effect. The cross-section varies  little with energy, especially as compared to, e.g., resonant scattering, and only the total column of electrons, bound and free, affect a propagating gamma-ray, so ionization and excitation are irrelevant. Therefore, the line opacity is constant throughout the ejecta for SN~Ia with approximately constant Y$_{e}$.

It is simplest to work in velocity space, which  then also provides gamma-ray line profiles of interest for detectors with sufficient energy resolution. Consider a geometrically thin spherical shell of expansion speed $v=r/t$, with radius r at time t since explosion, isotropically emitting gamma-ray photons. As is well-known, in the optically thin limit a distant observer will see a flat-topped line profile, with differential flux $f_{v}=F/2v$ from observed radial velocity  $v_{r}=-v$ to $v_{r}=+v$, where $F$ is the total line flux. The emission at $v_{r}=-v \, cos(\theta)$ comes from that circle on the sphere at angle $\theta$ between the line-of-sight and the radial line to the circle. We begin with this as the source spectrum, which includes the solid angle effects. 

We now consider a larger concentric spherical shell of finite thickness enclosing the emitting spherical surface. It has inner speed $v_{1}$ and outer speed $v_{2}=v_{1}+\Delta v$, where both speeds here are greater than $v$. This shell will attenuate photons from the source shell at all observed $v_{r}$, with the minimum attenuation for photons traveling radially at $v_{r}=-v$, The radial optical depth of the shell is $\tau_{r}=\kappa \rho \Delta v t$, where $\kappa$ is the gamma-ray opacity (e.g., 0.069 cm$^{2}$ g$^{-1}$ at 847 keV for matter with Y$_{e}$=0.5) and $\rho$ is the mass density of the shell. Photons from larger $v_{r}$ traverse a longer path through the shell, equal to one-half the difference of the two chords of the outer and inner surfaces of the shell, at impact parameter corresponding to that $v_{r}$. The path length traversed in velocity as a function of radial velocity is
\begin{equation}
  p(v_{r}) = (v_{2}^{2} - (v^{2}  - v_{r}^{2}))^\frac{1}{2} - (v_{1}^{2} - (v^{2}  - v_{r}^{2}))^\frac{1}{2}
\end{equation}
Then the transmission fraction of photons from this source shell through the outer shell is 
\begin{equation}\label{bigT}
T(v_{r}) = \exp(-\tau_{r} \, p(v_{r})/\Delta v).
\end{equation}
This transmission and therefore the resulting line profile after passing a single outer shell, has a ``V'' shaped profile. For multiple shells outside a single shell of emission, the emergent line shape is simply the flat profile times the products of the transmissions of all outer shells. 

Spherical shells interior to the emitting sphere scatter photons only from the back, receding part of the emitter. We label the inner shell properties as above, except $v_{1} < v_{2} < v$ now. Photons emitted close to the limb of the inner shell pass through the full chord of the sphere of radius $v_{2}$, while those passing close to the center of the shell traverse front and back parts of the shell, i.e., the full difference of the chords of inner and outer spheres. The path lengths in velocity through the shells are then
\begin{eqnarray}
 p(v_{r})   = & 2(v_{2}^{2} - (v^{2}  - v_{r}^{2}))^\frac{1}{2} - 
 2(v_{1}^{2} - (v^{2}  - v_{r}^{2}))^\frac{1}{2} & \ \  
 {\rm for}  \   v_{r}  > (v^{2} - v_{1}^{2})^\frac{1}{2}  \nonumber \\
  =  & 2(v_{2}^{2} - (v^{2}  - v_{r}^{2}))^\frac{1}{2} & \ \ 
 {\rm for} \  (v^{2} - v_{2}^{2})^\frac{1}{2} <  v_{r}  < (v^{2} - v_{1}^{2})^\frac{1}{2} .
\end{eqnarray}
The transmission through this inner shell is given by Eq.~\ref{bigT}, except T=1.0 for $v_{r} < (v^{2} - v_{2}^{2})^\frac{1}{2}$. The resulting line profile takes the shape of ``J'' on the high velocity side, and is flat on the approaching side. Again, the total transmission through multiple inner shells is the product of the individual shell transmissions at each radial velocity. 
\begin{figure}[ht!]
\plotone{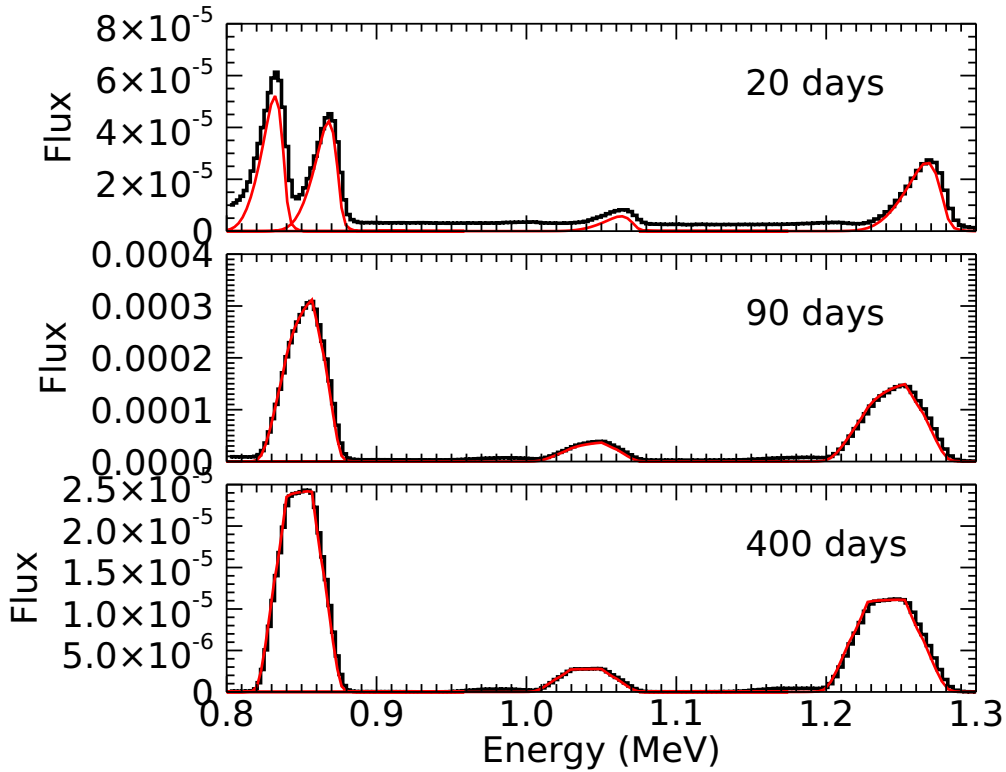}
\caption{Comparison of our analytic gamma-ray line escape calculation for individual lines from $^{56}$Ni and $^{56}$Co (red smooth curves) with Monte Carlo calculation spectra (black histograms) for model 7p0z22d20\_27 at three times. At 20 days, significant continuum from scattered photons is present, but the line profiles and fluxes agree well. The curves are slightly offset for presentation. \label{figappend}}
\end{figure}

A third case is when the emitting sphere is inside the scattering shell, with $v_{1} < v < v_{2}$.  This is effectively a combination of the above two cases, with shell edge speeds $v_{1}=v$ for the outer part, and $v_{2}=v$ for the inner part. The resulting path lengths through the shell are
\begin{eqnarray}
 p(v_{r})   = & (v_{2}^{2} - (v^{2}  - v_{r}^{2}))^\frac{1}{2} + v_{r}
  & \ \   
 {\rm for} \  -v < v_{r} < (v^{2} - v_{1}^{2})^\frac{1}{2}  \nonumber \\
  =  & (v_{2}^{2} - (v^{2}  - v_{r}^{2}))^\frac{1}{2} + v_{r} - 2(v_{1}^{2} - (v^{2}  - v_{r}^{2}))^\frac{1}{2} & \ \  
 {\rm for} \  (v^{2} - v_{1}^{2})^\frac{1}{2} < v_{r} < v .
\end{eqnarray}
The transmission through this one shell is again given by Eq.~\ref{bigT} with this path length. 
While this gives the correct escape for a thin emitting spherical surface, if the emission is spread throughout the scattering shell, we can subdivide this shell into many to improve the accuracy of the calculation. 

For an emitting shell, one multiplies its rectangular profile by the products of the transmissions of all shells. Summing these escaping profiles over all emitting shells gives the line flux profile for the entire object. If a supernova is modeled as a number spherical shells in homologous expansion, one has to specify only the source line luminosity of each shell, for example, the initial $^{56}$Ni mass, and the mass density of each shell, and the escaping fluxes and profiles of all gamma-ray lines are given for all times, without further approximations. Opacities for different lines are given by the ratios of the total Klein-Nishina cross-sections, and as all densities fall as t$^{-3}$, optical depths of all paths fall as t$^{-2}$. The scattered photons in the Compton continuum, which do carry significant information, are ignored here.
\begin{figure}[ht!]
\plotone{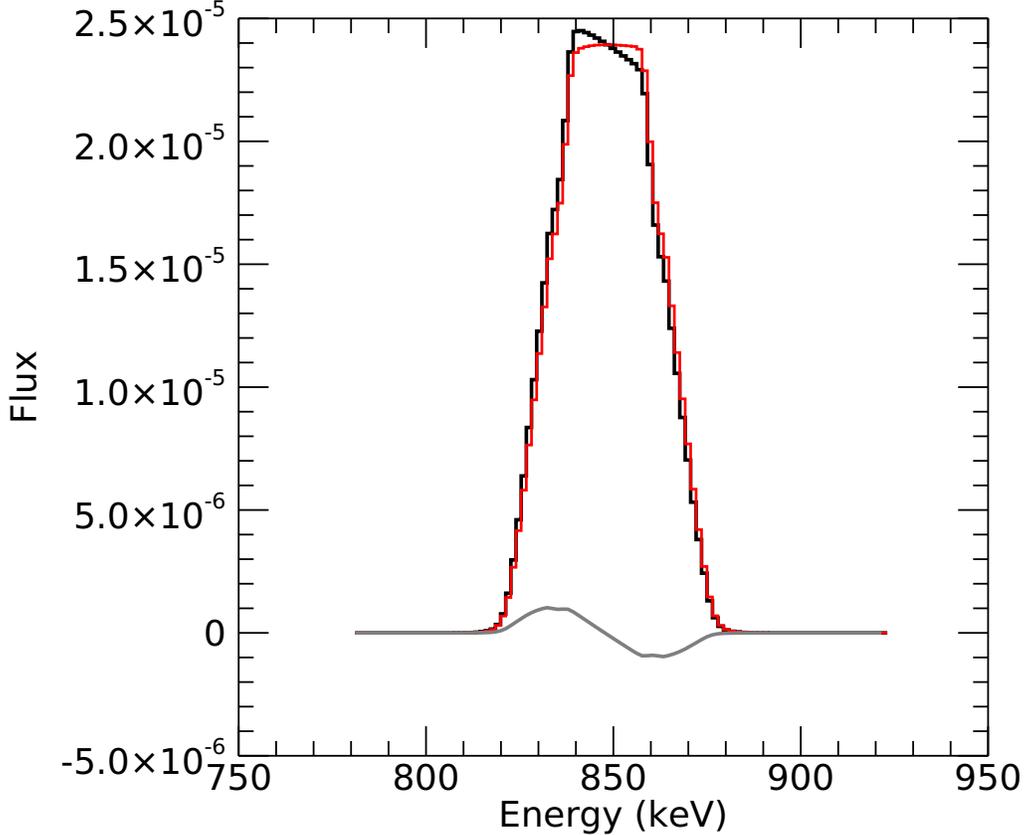}
\caption{Comparison of the 847 keV line profile with (black) and without (red) the light travel delays across the ejecta included, for 
 model 7p0z22d20\_27 at 400 days. The difference is shown as the gray curve. This is the largest effect of the times and lines we consider. \label{figdelay}}
\end{figure}

As an illustration, we take the zone velocities, mass densities, and $^{56}$Ni mass fractions of model 7p0z22d20\_27 \citep{2015ApJ...806..107D} and perform this calculation for the escaping line photons. We compare these to Monte Carlo calculations of the gamma-ray transport  for this model (Peter H\"oflich, private communication) and find good agreement for the lines at all times. In Figure~\ref{figappend} we show these comparisons at three times for the energy range that includes the $^{56}$Ni 812 keV line and the $^{56}$Co 847, 1038, and 1238 keV lines. At 20 days, the ejecta are still quite thick, and the scattered continuum is apparent in the Monte Carlo calculation. This can be separated from the lines in spectral analysis as we describe above. At later times, the agreement is excellent.

 In this method, the effect of light travel delays across the ejecta is simple to include, as it depends on radial velocity. To see the effect, we show the 847 keV line profile in Figure~\ref{figdelay} with and without delays included. It becomes important at times long relative to the decay lifetimes, so we illustrate this at 400 days for the $^{56}$Co line. Earlier, the effect is smaller, even for $^{56}$Ni lines,  which are seen mostly from the approaching side of the ejecta only. In this paper, we do not include this effect for the simulated data, because it was not included in the Monte Carlo calculations used to generate the data. However, it should be included in analyses of real supernovae, especially for high precision line profile measurements. In line flux measurements, the odd symmetry of the difference tends to cancel its effect at late times. 

Of course, Monte Carlo methods for this problem are accurate and have long been used to calculation emergent spectra \citep[e.g.,][]{2004ApJ...613.1101M}, as well as energy deposition, etc. However, if we want to determine density and radioactivity abundance profiles from gamma-ray data, we might have to evaluate the emergent spectra many times, for MCMC fitting perhaps tens of thousands,  to analyze a single data set, so a very fast, accurate method is preferable. Instead of summing the lines from different emitting shells, one can keep them separate, as they will have different fluxes and energy profile variations with time, and derive the distribution of initial $^{56}$Ni and total mass within the ejecta. Although we do not use it here, the zones where the gamma rays scatter are tracked, so lowest order energy deposition by gamma rays can be calculated very rapidly, and improved with simple  modifications.

\bibliography{../main}  
\bibliographystyle{aasjournal}      



\end{document}